\documentclass[journal]{IEEEtran}
\usepackage{cite}

\ifCLASSINFOpdf
\else
  \usepackage[dvips]{graphicx}
  \graphicspath{{./eps_submission/}}
  \DeclareGraphicsExtensions{.eps}
\fi
\usepackage[cmex10]{amsmath}
\usepackage{amsfonts}
\usepackage{cases}
\usepackage[font=footnotesize, caption=false]{subfig}
\usepackage{fixltx2e}

\ifCLASSOPTIONcaptionsoff
 \usepackage[nomarkers]{endfloat}
 \let\MYoriglatexcaption\caption
 \renewcommand{\caption}[2][\relax]{\MYoriglatexcaption[#2]{#2}}
\fi
 \let\MYorigsubfloat\subfloat
 \renewcommand{\subfloat}[2][\relax]{\MYorigsubfloat[]{#2}}
\usepackage{url}

\usepackage{booktabs}

\begin{document}
%
\title{Radioactive Decays in Geant4}
%
%
%
\author{Steffen~Hauf,
    Markus~Kuster,
	Matej~Bati\v{c},
	Zane~W.~Bell,
	Dieter~H.H.~Hoffmann,
	Philipp~M.~Lang,
	Stephan~Neff,
	Maria~Grazia~Pia,
	Georg~Weidenspointner
	and Andreas~Zoglauer
\thanks{Manuscript submitted January 28, 2012. This work has been supported by Deutsches Zentrum f\"{u}r Luft- und Raumfahrt e.V. (DLR) under grants 50 QR 0902 and 50 QR 1102.}
\thanks{S. Hauf and M. Kuster are with European XFEL GmbH, Hamburg, Germany (e-mail: steffen.hauf@xfel.eu)}
\thanks{D.H.H. Hoffmann, P.––M. Lang and S. Neff are with Institute for Nuclear Sciences, TU Darmstadt, Darmstadt, Germany}
\thanks{M.G. Pia and M. Bati\v{c} are with the INFN Genova, Genova, Italy}
\thanks{G. Weidenspointner is with the Max-Planck Halbleiter Labor, Munich, Germany and the Max-Planck Institut f{\"u}r extraterrestrische Physik, Garching, Germany}
\thanks{A. Zoglauer is with the Space Science Laboratory, University of California, Berkeley, CA, USA}%
\thanks{Z.W. Bell is with the Oak Ridge National Laboratory, Oak Ridge, TN, USA}}%
\maketitle

\begin{abstract}
The simulation of radioactive decays is a common task in Monte-Carlo systems such as Geant4. Usually, a system either uses an approach focusing on the simulations of every individual decay or an approach which simulates a large number of decays with a focus on correct overall statistics. The radioactive decay package presented in this work permits, for the first time, the use of both methods within the same simulation framework --- Geant4. 
The accuracy of the statistical approach in our new package, RDM-extended, and that of the existing Geant4 per-decay implementation (original RDM), which has also been refactored, are verified against the ENSDF database. The new verified package is beneficial for a wide range of experimental scenarios, as it enables researchers to choose the most appropriate approach for their Geant4-based application.
\end{abstract}


%
\IEEEpeerreviewmaketitle

\section{Introduction}

Radioactive decays and the resulting radiation play an important role for many experiments, either as an observable, as a background source, or even as a potential hazard when they are a source of radiation-induced damage for hardware and human beings. Detailed knowledge of the radiation inside and around an experiment and its detectors is thus required for a successful outcome of the experiment and to guarantee the safety of the operator. The increasing complexity of experiments often makes it prohibitively expensive, if not impossible, to completely determine the radiation characteristics and response of an experiment from measurements alone. In order to circumvent these limitations, it has become increasingly important to estimate an experiment's radiation and response characteristics with the help of computer simulations. 

General-purpose Monte-Carlo simulation codes either focus on the correct simulation of individual decays (e.g., Geant4~\cite{Agostinelli2003250,
 2006ITNS...53..270A}, see Section~\ref{sec:problem_domain}) or the statistical outcome of many decays (e.g., MCNP~\cite{MCNP, MCNPX} and FLUKA~\cite{2001amcr.conf..955F}).
 
 Whereas the first approach may be inefficient if the individual decay is not of interest, the latter approach does not allow for the physically correct simulation of an individual decay and its associated effects. General purpose Monte Carlo codes would benefit from the capability of providing both approaches in the same environment, in response to the simulation requirements of different experimental scenarios. 

In the following a software package for the simulation of radioactive decay, which realizes both approaches for Geant4, is presented. This package includes a refactored implementation of the existing Geant4 per-decay approach~\cite{TruscottG4}, and extends the functionality of Geant4 radioactive decay simulation by a novel implementation based on a statistical approach. It is based on the ENSDF (Evaluated Nuclear Structure Data File) data library~\cite{ENSDF}, which was chosen due to its widespread usage in the nuclear science community.

This paper reports on the verification of both implemented approaches against a large set of evaluated data. To the best of the authors' knowledge, such a thorough verification of Geant4 radioactive decay simulation has not yet been documented in the literature.
The experimental validation of the software for both approaches is reported on in a separate paper~\cite{RadDecay2012_2}.

\section{Radioactive Decay Physics}
\label{sec:physics}
Radioactive decay is a physical process where an atomic nucleus of an unstable atom transmutes into a lower energy state by spontaneous emission of ionizing radiation. The process does not require external interactions to occur. It results from either nucleus-internal processes or interactions of the nucleus with (inner) shell electrons. A brief overview of the main physics of radioactive decay is summarized here to facilitate the comprehension of the functionality of the software described in this paper. 

Different types of radioactive decay are commonly identified according to the type of emitted particles.
\begin{itemize}
\item During an $\mathrm{\alpha}$-decay a He--nucleus is emitted from the parent nucleus. This results in a daughter nucleus with two fewer protons and two fewer neutrons than the parent nucleus. 

\item The $\mathrm{\beta^{-}}$-decay is a weak process during which a neutron is converted into a proton. An electron and anti-neutrino are emitted by the parent nucleus: consequently, the atomic number of the daughter nucleus increases by one and the atomic mass number stays constant. The electron and anti-neutrino share the energy released during the decay. Since both particles are not bound in their final state, their energy distribution follows a continuous spectrum.

\item During a $\mathrm{\beta^{+}}$-decay a bound proton of a nucleus is converted into a neutron. A positron and a neutrino are emitted by the parent nucleus; the atomic number decreases by one and the atomic mass number stays constant. Similar to $\mathrm{\beta^{-}}$-decays, both particles are not bound in their final state, and accordingly, their energy distribution follows a continuous spectrum.

\item If a daughter nucleus is left in an excited state, after a transmutation by the previously mentioned decay types, it can deexcite by emitting $\mathrm{\gamma}$-radiation. In case the excited daughter state is a long-lived (metastable) state, its deexcitation is called isomeric transition, which will also result in $\mathrm{\gamma}$-radiation. In both cases the atomic number and atomic mass number remain unchanged.

\item During an electron capture, the parent nucleus absorbs an inner shell electron (usually K- and L-shell electrons) and simultaneously emits a neutrino. During this process, which is also called inverse $\mathrm{\beta}$-decay, a proton is transmuted into a neutron, thus the atomic number decreases by one and the atomic mass number stays constant. In contrast to a $\mathrm{\beta}$-decay, an electron capture is a two-body decay, resulting in a discrete neutrino energy. 

\item As an alternative process to $\mathrm{\gamma}$-emission, an excited nucleus can return to its ground state by transferring its excitation energy to one of the lower shell electrons of the atom. This process is called internal conversion, and results in the emission of an electron by the atom, leaving the atom in an excited state. The electron carries a discrete fraction of the decay energy, and by this is distinct from $\mathrm{\beta}$-particles with continuous energy spectra. As with $\mathrm{\gamma}$-decays, no transmutation of the nucleus takes place, and both the atomic number and atomic mass number remain unchanged.

\end{itemize}

Radioactive decay is a stochastic process. The time at which a given unstable atom decays is not predetermined; instead decays occur with a certain probability. In consequence, experiments will measure statistical observables such as the amount of ionizing radiation of a certain type and energy emitted within a given time period. 

Due to practical impossibility of calculating all relevant parameters from theory, the simulation of radioactive decay physics for a large number of decays or decay chains relies on the usage of empirical or pre-calculated data.

\section{Experimental Scenarios}
\label{sec:exp_req}
Many experiments measure the time-accumulated statistical distributions of observables such as energy, type, momentum and timing of radiation resulting from radioactive decays. Often the radioactive decay products are not the intended observable but contribute to the experiment's or application's measured data as background. 

Radioactive decay physics modeling plays an important role in various fields; the overview summarized here is not intended to be exhaustive.

Measurement and analysis of the properties of decay chains of naturally abundant radioactive isotopes is commonly performed in material sciences, radiation safety and nuclear proliferation monitoring. Radioactive decay modeling is also of interest to experimental scenarios involving the measurement of material properties after irradiation: for instance, of a sample irradiated by a neutron beam.

The study of activation and build-up of radioactive nuclei is relevant to various applicative domains: nuclear reactors (fission and fusion), particle accelerators or intense light sources, where the statistical effect of many decays and activations is relevant. It also concerns space-borne X-ray and $\mathrm{\gamma}$-ray instruments: in these scenarios meta-stable states usually must be accounted for, and the statistical effect of many decays and activations is relevant. Additionally, individual decays contribute to the prompt instrument background. The estimation of the in-orbit cosmic-ray induced background is important for space-based detectors to distinguish individual radioactive emission from intended observables.

Low background astro-particle physics experiments are concerned with the influence of the cosmic-ray induced background and natural radioactivity: here accurate simulation of the individual decays' spatial and temporal distribution can be important.

\section{Foundations for a Radioactive Decay Simulation}
\subsection{Requirements}
\label{sec:problem_domain}

The simulation of radioactive decays consists of the task of decaying an unstable nucleus and generating the resulting products.
 
The decay of the parent nucleus should proceed according to the physical parameters governing it: decay type, initial excitation and half-life time. A daughter nucleus should be produced, with the physical properties resulting from the decay:  atomic number and mass determined by the decay type, excitation energy and kinetics determined by parent kinetics and decay kinetics. Secondary particles and radiation associated with the decay (e.g. $\mathrm{\beta}$-, $\mathrm{\gamma}$- and $\mathrm{\alpha}$-emission, neutrinos and conversion electrons) should be generated.

The software should handle the deexcitation of the daughter atom, involving the production of $\mathrm{\gamma}$- and X-rays, and of Auger-electrons.

Theoretical calculations of the required parameters are not practically feasible in the course of the simulation; therefore the algorithm must use empirical or pre-calculated data.

Radioactive decays of nuclei are often associated with prior activation of stable nuclei of a given material; examples for such applications include shielding analysis and radiation safety analysis. This scenario requires the ability of dealing with activation, and thus replenishment of nuclei, within the simulation. Meta-stable states and isomeric transitions should be taken into account. 

In addition to these functional~\cite{6146379}, physics-induced requirements, the software should take into account non-functional ones, which derive from the experimental context of the simulation: the possibility to efficiently simulate a large number of decays, when the physical accuracy of the individual decay is of less importance, but overall statistics are relevant; the possibility of efficiently simulating decay chains, when intermediate products in a chain are of lesser interest; the provision of user input.

If radioactive decay simulation occurs in the context of a more general Monte Carlo simulation system, the software responsible for the radioactive decay process should interact with other components of the system.

\subsection{Problem Domain Analysis}
\label{sec:probl-doma-analys}
Software objects with specialized responsibilities collaborate to satisfy the requirements mentioned in Section~\ref{sec:problem_domain}. For the code
presented in this work the division of responsibilities is as
follows:
\begin{itemize}
	\item data management
	\item sampling of the (discrete) emission resulting from the
     individual decay and generation of the daughter nucleus
	\item calculation of the $\mathrm{\beta}$-emission spectrum,
	\item calculation of the number of nuclei within a decay chain
     --- which may include activation
	\item the user interface
	\item the interface with the Monte-Carlo code
\end{itemize}

\section{The ENSDF Data}
The sampling of radioactive decays will usually rely on the use of empirical or pre-calculated data. The Evaluated Nuclear Structure Datafile (ENSDF)~\cite{ENSDF} is one such collection of data, which for instance contains information on half-life times, decay types, branching ratios, emission energies and transition types. It is maintained by the National Nuclear Data Center (NNDC) and distributed, amongst others, by the International Atomic Energy Agency (IAEA). 

ENSDF is an evaluated library, i.e. it contains data from experiments and theoretical calculations which are recommended for use as a reference after a critical analysis of uncertainties, inter- and extrapolation methods and underlying models has been performed. IAEA defines it as the master library for evaluated experimental structure and decay data~\cite{ensdf_iaea}. Other specialized libraries and bibliographic databases, such as NuDAT~\cite{sonzogni:574}, CINDA~\cite{cinda} and NSR~\cite{Pritychenko2011213}, exist as well, but are commonly derived from or related to ENSDF. Therefore ENSDF is frequently 
considered the de-facto standard for nuclear structure and radiation data.

For $\mathrm{\gamma}$-ray intensities and energies the ENSDF data usually consists of evaluated measurements. Conversion electron
intensities and energies are derived from theory.

However, the data present in ENSDF are not sufficient to calculate all atomic deexcitation emissions. Specifically, data on electron binding energies, fluorescence and Auger-electron yields are necessary to calculate the respective intensities. To mitigate this problem, analysis programs distributed with ENSDF, such as {\tt RADLST}~\cite{radlist}, are supplied with the necessary information taken from data of Bearden and Burr~\cite{RevModPhys.39.125} and Bambynek~\textit{et al.}~\cite{1972RvMP...44..716B}. In order to stay consistent with these ENSDF-related programs, the aforementioned data are also used to derive quantities for the radioactive decay database of the RDM-extended package and the verification data used in this work.

\section{Radioactive Decay in Monte Carlo Codes}
\label{sec:MCCodes}
Models for the simulation of radioactive decays exist in most
general-purpose Monte-Carlo codes. 

MCNP(X)~\cite{MCNP, MCNPX} does not include a full radioactive
 decay simulation by default except for the generation of delayed
 $\mathrm{\gamma}$-rays resulting from decays sampled from MCNP's photon data library (phtlib). Instead MCNP can be linked via scripts to specialized
 codes such as ORIGEN2~\cite{Croff:Origen2} or
 CINDER90~\cite{CINDER90, 2005AIPC..769..195G}. Both of these use their own
 data libraries. These codes are generally used to model reactor fuel
 cycles and accelerator induced transmutations, but also provide
 functionality for simulating radioactive decays and decay
 chains. Due to the codes' nature, they include replenishment through
 activation. The $\mathrm{\alpha}$- and $\mathrm{\beta}$-emission needs
 to be generated from tabulated user input. As a result, MCNP is
 specialized on the simulation of many decays and the resulting
 statistics.

FLUKA~\cite{2001amcr.conf..955F} generates and transports
 $\mathrm{\beta}$- and $\mathrm{\gamma}$-radiation, but only started including
 $\mathrm{\alpha}$-radiation in its latest release. Decay chains are possible and include
 replenishment through activation. FLUKA uses its own data libraries,
 largely based on NNDC (National Nuclear Data Center) data and thus ENSDF. Similar
 to MCNP, the emphasis lies on the simulation of a large number of
 decays.

In Geant4~\cite{Agostinelli2003250, 2006ITNS...53..270A}
 radioactive decays are treated on a per-decay level, based on data taken
 from ENSDF~\cite{ENSDF}. Decay chains including activation are
 possible and produce the associated decay
 emission. The $\mathrm{\alpha}$- and $\mathrm{\beta}$-emission are
 sampled from the decay database. Deexcitation radiation of the
 daughter nucleus is not produced by the radioactive decay simulation
 itself, but by other physics processes included in Geant4, which use
 their respective databases. The emphasis lies on the per-decay
 simulation, not on the sampling of a large number of decays.

None of the above mentioned Monte-Carlo codes allows the simulation of either individual decays or a statistical treatment of many decays in the same software environment.

\section{Radioactive Decay in Geant4}
\label{sec:geant4_decay}
A package for the simulation of radioactive decays~\cite{TruscottG4},~\cite{896281} has been available in Geant4 since version 2.0, where it was named {\it radiative\_decay}. Since Geant4 version 6.0 it has been named {\it radioactive\_decay}, although it is conventionally known as the Geant4 RDM (Radioactive Decay Module). This code was originally developed by P. Truscott and F. Lei; it implements per-decay sampling.

The following discussion is based on the radioactive decay
code of Geant4 9.4p04, but is also pertinent to subsequent versions 9.5 and 9.6, the latter one being current at the time of submission of this paper. In these latter versions the problem of not producing fluorescence emission in case of decays other than electron capture has been addressed and the handling of forbidden $\mathrm{\beta}$-decays has been added, but all other features and problems of the implementation mentioned in this work are still valid.

\begin{figure*}
\centerline{\includegraphics[width=6.5in]{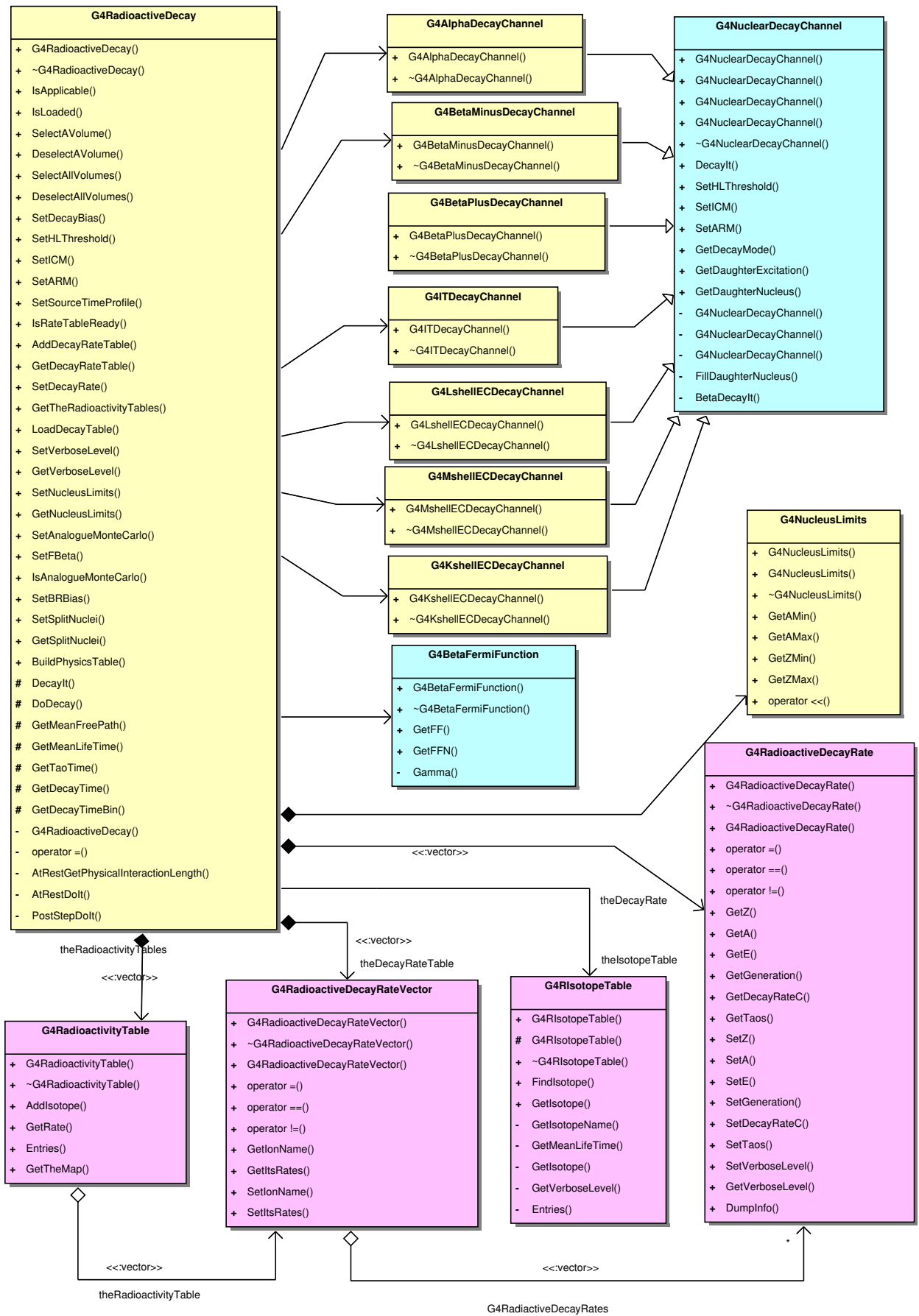}%
}
\caption{A UML class diagram of the original RDM, which implements per-decay sampling,
 showing the relations between the individual classes. The colors (grey shades)
 reflect the responsibilities and are matched to the activity diagram
 in Fig.~\ref{fig:activity_old}: physics simulation (blue/medium grey) and
 decay chain processing (purple/dark grey). Yellow/light grey objects can not be attributed to a single responsibility. Data retrieval, initialization and
 management is provided by the main class ({\it
  G4RadioactiveDecay}). The individual decay channels initialize a
 {\it G4NuclearDecayChannel} object which includes the physics of all
 decay types in one class.}
\label{fig:class_old}
\end{figure*}

\begin{figure*}
\centerline{\includegraphics[width=7.in]{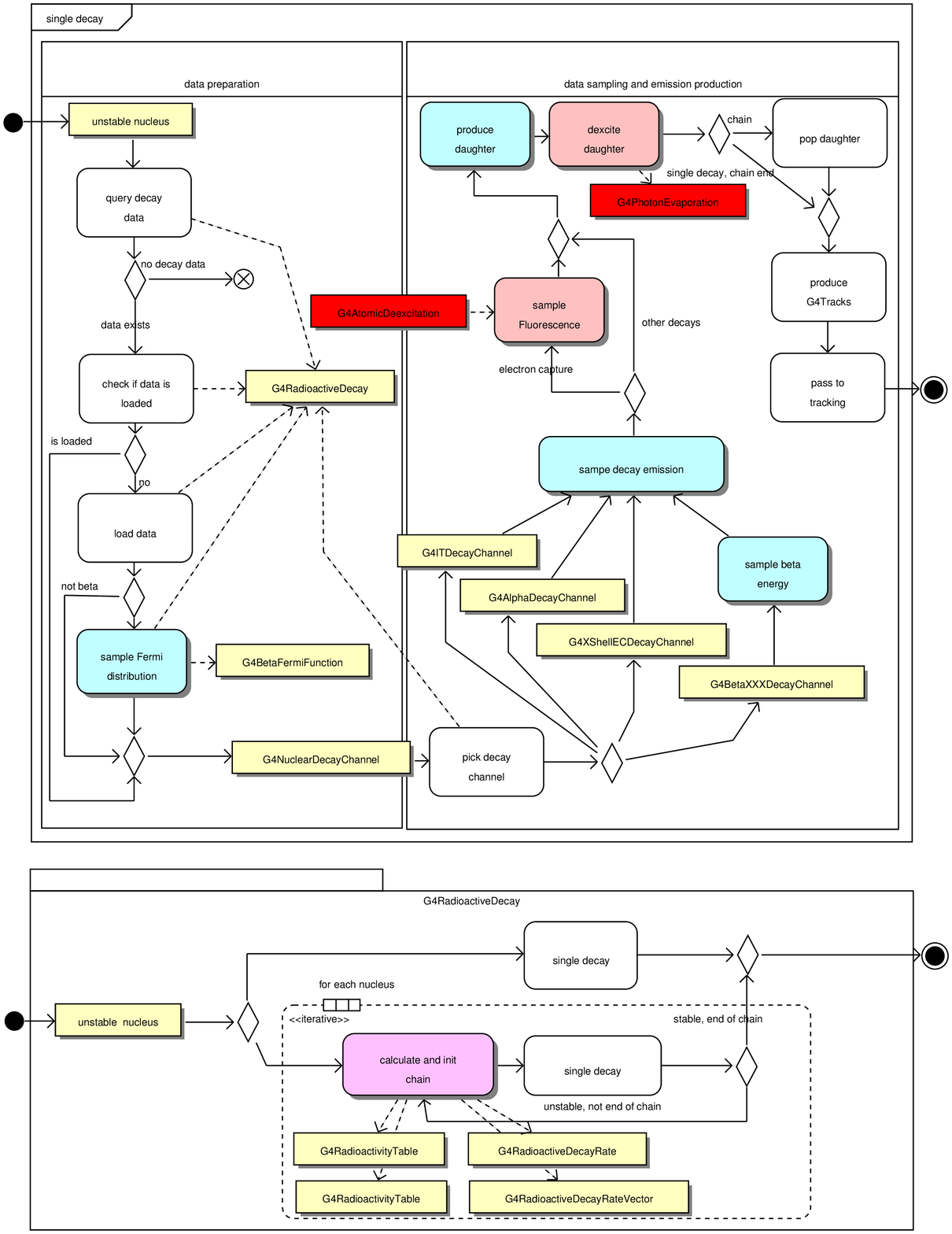}%
}
\caption{A UML activity diagram of the original RDM, which implements per-decay sampling
 showing the workflow of the code when decaying a nucleus. The color (grey shade) scheme is
 matched to the class diagram in Fig.~\ref{fig:class_old} and
 reflects different activity types: physics simulation within the
 radioactive decay code (blue/medium grey), physics simulation by other processes
 (red/darkest grey) and decay chain calculations (purple/dark grey). Yellow/light grey objects can not be attributed to a single responsibility.}
\label{fig:activity_old}
\end{figure*}

A Unified Modeling Language (UML)\cite{Rumbaugh2004} class diagram of the RDM code in Geant4 9.4p04 is shown in
Fig.~\ref{fig:class_old}. It shows the cooperation between the different classes in the code. This diagram is supplemented by the
activity diagram shown in Fig.~\ref{fig:activity_old}. The two diagrams highlight problems~\cite{Fowler:424198} inherent in the code's design: 
\begin{itemize}
\item Since each decay type is defined by distinctive physics, it would make sense to implement the individual types as separate objects. Instead, in the original Geant4 RDM package all decay types are implemented together in the {\it G4NuclearDecayChannel} class. The decay type classes merely provide an interface to this class with decay type-dependent initialization parameters. This complicates unit tests of individual components. Additionally, a much larger amount of code has to be checked if, for instance, an error is found for one decay type.
\item Whereas the decay physics for each type is distinct, the interface to each type is similar: all decay type objects should have a method to produce decay emission and a daughter nucleus. Such an interface could be provided by a common base class. In the existing design the {\it G4RadioactiveDecay} class needs to know the implementation and interface details of each individual decay type.
\item Objects should have one specific responsibility, e.g. the interface to the data library. This is not the case: the {\it G4RadioactiveDecay} class is responsible for initializing and loading values from the data libraries, initializing the decay types and the variance reduction and decay chain handling. Such a design again enlarges the fraction of code which needs to be checked for a specific error or maintained for the update of a specific responsibility. Additionally, due to inadequate domain decomposition, two distinct responsibilities –- the simulation of radioactive decay and event biasing -– are mixed in the same class.
\item In case of the $\mathrm{\beta}$-Fermi-function implementation the interface is implementation-dependent. Changing the algorithm may thus also involve changing the interface --- and as a result all other code parts depending on this class. 
\end{itemize}

A new package, named RDM-extended, has been developed to address existing issues of Geant4 RDM, and to extend and improve the capabilities of radioactive decay simulation in a Geant4-based environment.

The software design of the RDM-extended package follows an object-oriented programming approach with clear responsibility definitions. For this design the relevant entities and requirements for radioactive decay physics have been identified. Decay type-dependent approaches have been considered alongside common tasks for all decay types. The design also takes into account that two sampling methods are to be handled within a common framework, and that the external classes the code depends on may be subject to interface changes.

\begin{figure*}
\centerline{\includegraphics[width=7.in]{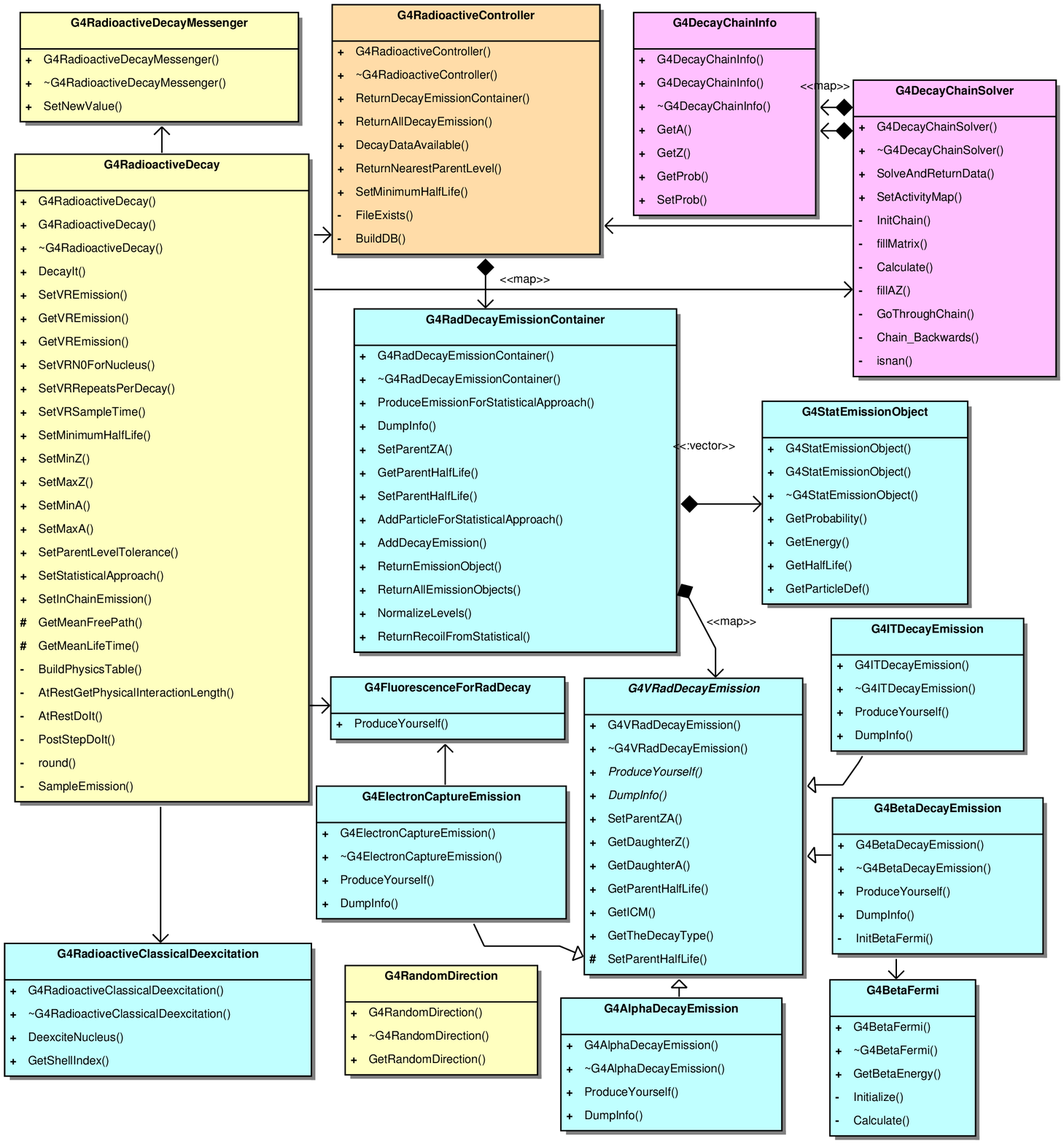}%
}
\caption{A UML class diagram for the RDM-extended package implementing a refactored per-decay sampling and a statistical sampling method. The colors (grey shades) reflect the
 responsibilities and are matched to the activity diagram in
 Fig.~\ref{fig:activity_new}: physics simulation (blue/medium grey), data
 management and preparation (orange/medium dark grey) and decay chain processing
 (purple/dark grey). Each decay type class is responsible for simulating the type's
 characteristic physics. The deexcitation and fluorescence emission
 for the statistical simulation approach is produced by the
 {\it G4VDecayEmission} base class.}
\label{fig:class_new}
\end{figure*}

\begin{figure*}
\centerline{\includegraphics[width=7.in]{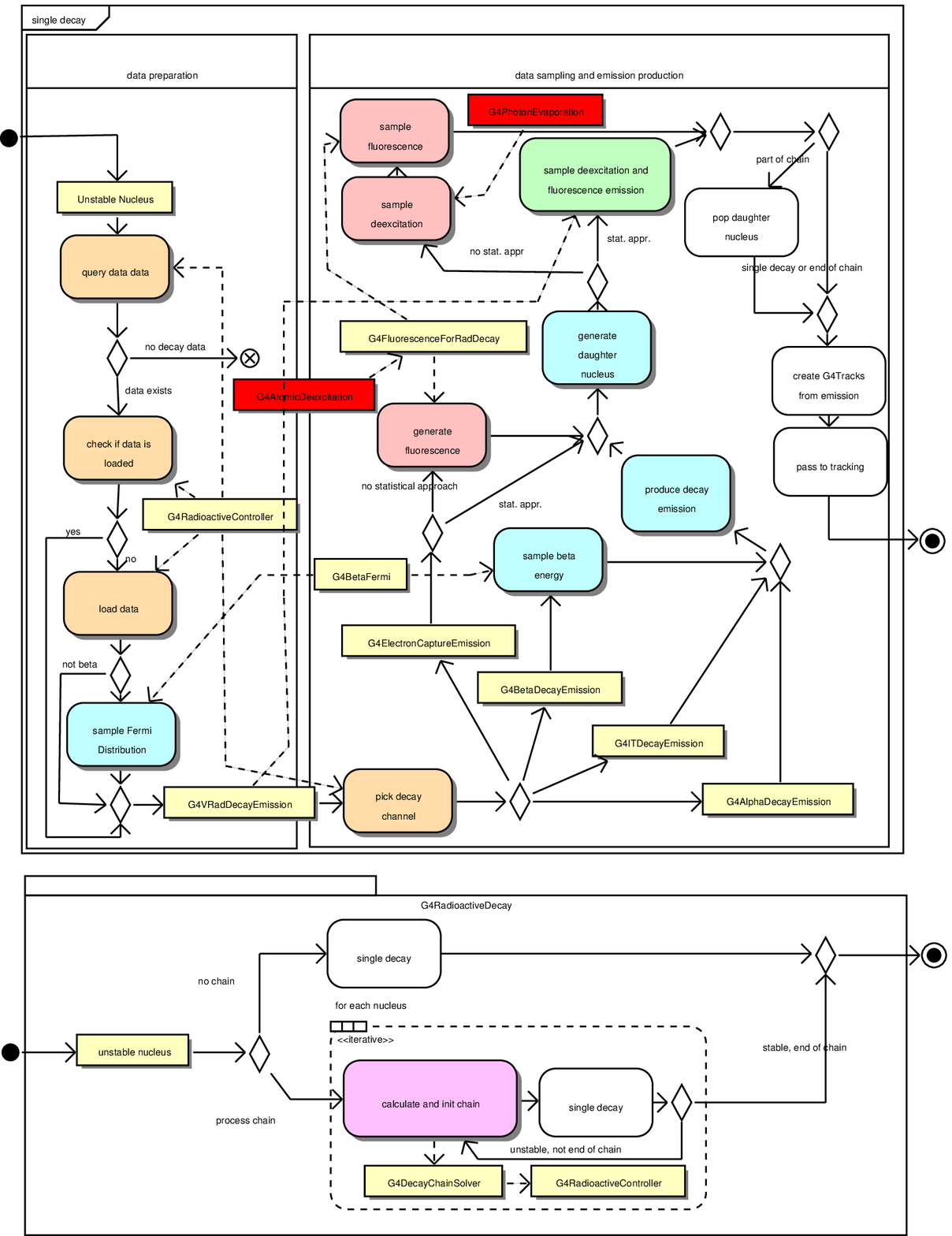}%
}
\caption{A UML activity diagram for the RDM-extended package implementing a refactored per-decay sampling and a statistical sampling method, showing the
  workflow of the code when decaying a nucleus (top) and handling of decay chains (bottom). The color (grey shade) scheme is matched to
 the class diagram in Fig.~\ref{fig:class_new} and reflects
 different activity types: physics simulation within the radioactive
 decay code(blue/medium grey), physics simulation by other processes (red/darkest grey), data
 management and preparation (orange/medium dark grey) and decay chain calculations
 (purple/dark grey). The emission produced by the statistical decay approach is
 highlighted in green/medium light grey.}
\label{fig:activity_new}
\end{figure*}

The UML diagrams in Fig.~\ref{fig:class_new} and~\ref{fig:activity_new} document that the responsibilities of objects are clearly defined and that functionality is neither duplicated nor aggregated into non-specialized classes.

An example for implementing functionality only once is the class {\it G4RandomDirection}, which provides functionality for sampling random particle momentum vectors to the classes modeling the individual decay types and the statistical sampling.

The {\it G4RadioactiveDecay} class of the original RDM (Fig.~\ref{fig:class_old}) is an example of a non-specialized class: it is responsible for physics simulation, data preparation and decay chain handling. In contrast the {\it G4RadioactiveDecay} class of the RDM-extended is a pure management class, which coordinates the interaction of specialized classes for the aforementioned tasks: the different emission classes (physics simulation), {\it G4RadioactiveController} (data-management) and {\it G4DecayChainSolver} (decay chains).

The RDM-extended package also respects encapsulation rules. Furthermore, physics functionality implemented in the different classes can be combined as needed, thus allowing both sampling approaches to use a common code base for functionality required by both.

The addition of a statistical sampling approach is reflected in the activity diagram: the initial configuration and library data access are common for both design approaches. They differ after a decay channel is chosen: in the per-decay approach fluorescence emission is sampled for electron capture decays by delegating responsibility to the {\it G4AtomicDeexcitation} class. At a later point the deexcitation emission of the nucleus is sampled by using the {\it G4PhotonEvaporation} class. If the statistical approach is chosen instead, all photon emission and discrete electron emission processes are sampled simultaneously at the end of handling decay physics.

\section{Per-Decay Sampling}
\label{sec:per_decay}
Per-decay sampling is already present in Geant4. This original RDM code provides the required functionality: therefore the general approaches to physical treatment and data handling were preserved, but refactored before their inclusion into the RDM-extended in order to conform to the design discussed in the previous section. This refactored per-decay sampling is consistent with the original RDM. Additionally, small errors in the physical treatment were addressed. As a result the structure of the code is different from the original one, but the inherent functionality is conserved.

Per-decay sampling is based on reprocessed ENSDF data both in the original RDM and the RDM-extended package. For the RDM-extended we reprocessed the data using a parser, which implements the ENSDF format as given in the ENSDF manual~\cite{ENSDF}. The
Geant4 classes handling nuclear and atomic deexcitation in the per-decay approach use their own libraries. For the {\it G4PhotonEvaporation}
class (nuclear deexcitation) this is ENSDF-based, with conversion
electron probabilities compiled from~\cite{Band1976433, Roesel197891,
 Hager19681}. The {\it G4AtomicDeexcitation}
class (atomic deexcitation), which has been previously
validated in~\cite{Pia4237413} and~\cite{pia2009validation}, uses EADL~\cite{Perkins:236347} data. Concerns about the accuracy of the EADL data have been reported in~\cite{pia2011evaluation}.

The electron capture probabilities included in both radioactive decay libraries are not given in ENSDF directly. They need to be calculated using an additional data source, which gives fluorescence yields and binding energy information. For consistency reasons the atomic data file distributed with ENSDF is used for the RDM-extended package, which is based on data by Bearden and Burr and Bambynek {\it et al.}~\cite{RevModPhys.39.125, 1972RvMP...44..716B}. This also facilitates comparisons with ENSDF-based online-databases such as NuDat, which use the same data. 

Further refactoring is foreseen to use a package for atomic data management exploiting the results of~\cite{2012JPhCS.396b2039B} and an improved package for the simulation of atomic relaxation, which are currently under development. These packages are intended to satisfy requirements common to the simulation of radioactive decay and of electromagnetic interactions.

The radioactive decay database of both codes is supplied in plain text on a per isotope basis. 
 
\subsection{Energy Conservation}
For a physically correct outcome energy needs to be conserved for
the decay, all decay emissions and when deexciting the daughter
nucleus. Nuclear deexcitation is handled by the {\it G4PhotonEvaporation}
class, which uses a reprocessed ENSDF-based data library different from the radioactive decay class. Since it is not guaranteed that the level-energies in the two data
libraries are exactly the same, a possible deviation has to be
taken into account. This is done in the original Geant4 RDM on three levels:

\begin{itemize}
\item If the level energy passed by the radioactive decay code does not correspond to a tabulated level in the {\it G4PhotonEvaporation} library, the nearest level present in the {\it G4PhotonEvaporation} database is used to retrieve the possible transitions. The initial deexcitation step occurs from the energy passed by the decay code and will transition to an energy tabulated in the deexcitation library. The energy of the first $\mathrm{\gamma}$-ray will thus not be in accordance with the tabulated evaporation data, but deviate by the difference between the tabulated radioactive decay and evaporation library values. Further transitions will then result in emission at energies in accordance with the {\it G4PhotonEvaporation} data. 

\item For all transitions along a deexcitation chain it is checked if
 the level resulting from the transition is within a tolerance of
 $1\,\mathrm{keV}$ of the ground state. If this is the case, the
 photon energy is set to the energy of the excited level and the
 nucleus is deexcited to the ground state. Again energy is conserved,
 but the energy of the emitted photon may not be in
 accordance with the tabulated data.

\item The final state energy is passed back to the radioactive decay code. In case this energy is not $0$, but below $1\,\mathrm{keV}$, the daughter's excitation energy is set to $0\,\mathrm{keV}$. Here energy is not conserved. Otherwise the decay code outputs an excited (possibly meta-stable) daughter nucleus as defined by the data library.
\end{itemize}
The first two conservation treatments are necessary if slightly
divergent data libraries exist. They are handled internally by the {\it G4PhotonEvaporation} class for all processes which delegate to it, and have thus not been
altered as part of the radioactive decay code development. Rectifying these divergencies would require a consolidation of
the Geant4 data libraries, which exceeds the scope of this work. The
last approach to energy conservation is physically unjustifiable and accordingly has been
corrected in the refactored implementation. In case of an energy mismatch, the energy of an excited nucleus is deposited in the geometrical volume it is located in. Whereas this
does not adress the library inconsistency, which leads to the energy
mismatch, it does conserve energy and is accordingly seen as the
preferable solution.

\subsection{Momentum Conservation and Recoil of Nucleus}
Like energy, momentum should be conserved during the complete decay simulation. In the per-decay approach nuclear deexcitation is delegated to {\it G4PhotonEvaporation}, which also generates the momenta of the $\mathrm{\gamma}$-rays and conversion-electrons in a deexcitation cascade. The radioactive decay simulation is then responsible to account for the resulting recoil of the daughter nucleus.

Similarly, the momenta of any fluorescence and Auger-emission are sampled by the {\it G4AtomicDeexcitation} class and should also be taken into account as vectorial quantities. This is not the case in the original Geant4 RDM code: here the difference of all produced emissions' summed kinetic energies and the binding energy of the innermost vacated shell are added to the kinetic energy of the daughter nucleus. Whereas the computation of scalar values might result in a marginal performance increase, it neglects that the emission is not unidirectional, but isotropic. A full vectorial 
treatment is the physically accurate solution and is thus implemented in the RDM-extended package.

\section{Statistical Sampling}
\label{sec:stat}
The statistical sampling approach discussed in this section is a novelty for Geant4-based radioactive decay simulation. The necessity for such an approach has become evident as a consequence of the experimental requirements
identified in Section~\ref{sec:exp_req}. Statistical sampling in this
context means that the full simulation of an individual
decay is considered of lesser importance as long as the emission
and nuclei produced by many decays on average lead to a physically
correct result.

Because all decay emission is treated as independent of each other and
the intensity of each emission is known, sampling generally reduces to
the problem of repeated random number generation within the intensity
range of $0\ldots 1$. Additionally, one has to take into account that intensities
$I_\mathrm{{em}}>1$ may exist, if both nuclear and atomic deexcitation result in radiation at the same energy. In the latter case at least
$I_\mathrm{{em}}\,\pmod 1$ emissions will be generated for every
decay.

This approach does not require to take allowed or forbidden level
transitions into account during nuclear deexcitation since the order of occurrence of transitions is considered irrelevant, as long as for a large enough number of transitions, each occurs with the correct probability, in result yielding the correct intensity. 

These intensities are included in ENSDF alongside normalization information and as such all required
decay information may be retrieved from a single consistent
library. Accordingly, energy mismatches between libraries do not occur and no interdependencies between different physics classes exist. As a
consequence, an alteration of tabulated library values will manifest
itself in a straight-forward and immediate fashion in the simulated decay intensities and energies.

The conservation of energy and momentum is simplified, as the
information of momenta and energy of all decay emissions is known to
the radioactive decay simulation. Accordingly, it can be taken into
account in a physically meaningful fashion. 

In consequence, statistical sampling is a more efficient, straight-forward, ENSDF-consistent and performant approach for simulating radioactive decays for the majority of experimental applications which do not require knowledge of individual decays. 

\section{Sampling-Method-Independent Functionality in the RDM-Extended Radioactive Decay Code}

In the previous two sections a distinction between the per-decay and statistical sampling approach was made. For a complete solution to radioactive decay physics, in accordance with the requirements, sampling schemes for $\mathrm{\beta}$-emission and decay chains are necessary. These are approach-independent as they occur in both sampling scenarios.

\subsection{$\mathrm{\beta}$-Fermi-Function - Sampling of the Continuous $\mathrm{\beta}$-Spectrum}
\label{sec:bf_em}
$\mathrm{\beta}$-decay is an unbound three-body decay and the emitted radiation will thus have a continuous spectrum. This spectrum can be sampled using the $\mathrm{\beta}$-Fermi-function, which also takes corrections resulting from the interaction of the charged nucleus with the $\mathrm{\beta}$-particle into account.

The parameters passed to the $\mathrm{\beta}$-Fermi function should be physics-relevant, but code-independent. In particular this means that, instead of passing a binning scheme dependent energy, as is done in the original Geant4 RDM implementation, the physical parameters $Z$ and endpoint energy $E_0$, as well as the decay type ($\mathrm{\beta^{-}}$ or $\mathrm{\beta^{+}}$) and optionally forbiddenness, are passed in the RDM-extended implementation. This parameter set is sufficient for many Fermi-function approximations, like to those summarized by Venkataramaiah~{\it et al.} in~\cite{Venkataramaiah}. Currently, the computationally most performant approximation given therein is used in the RDM-extended package for calculating the Fermi correction factor $F(Z,E)$ for an isotope with atomic number $Z$ with the total energy $E$(given in MeV):
\begin{equation}
 F(Z,E) = [A+B/(E-1)]^{\frac{1}{2}}.
\end{equation}
The constants $A$ and $B$ were determined by Venkataramaiah~{\it
 et al.} through linear regression using the data from
Rose~\cite{rose1955}, and were found to satisfy:

\begin{subnumcases}{A=}
	1+a_{0}\,\mathrm{exp}(b_{0}Z) & for $Z\ge16$ \\
	7.3\times 10^{-2}\, Z+9.4\times 10^{-1} & for $Z<16$
\end{subnumcases}
with 
\begin{eqnarray}
	\nonumber a_{0} & = & 404.56\times 10^{-3} \\
	\nonumber b_{0} & = & 73.184\times 10^{-3} 
\end{eqnarray}
and
\begin{equation}
	B = a\,Z\,\mathrm{exp}(bZ)
\end{equation}
with
\begin{subnumcases}{a=}
  \nonumber 5.5465\times 10^{-3} & for $Z\le56$ \\
  \nonumber 1.2277\times 10^{-3} & for $Z>56$
\end{subnumcases}
\begin{subnumcases}{b=}
  \nonumber 76.929\times10^{-3} & for $Z\le56$ \\
  \nonumber 101.22\times 10^{-3} & for $Z>56$.
\end{subnumcases}
The correction factor $F(Z,E)$ is then input into the $\mathrm{\beta}$-Fermi
function
\begin{equation}
 N(p)\,dp=F(Z,E)\, p^2 \,(E_0-E)^2 \,dp,
\end{equation}
where $E_0$ is the end-point energy of the $\mathrm{\beta}$-spectrum, obtained
from the tabulated radioactive decay data library, $E$ is the total
energy of a $\mathrm{\beta}$-particle with momentum $p$. If a given
parameter set is computed for the first time, a tabulated energy
distribution is calculated, from which the $\mathrm{\beta}$-particle
energy is drawn. Future occurrences of the same parameter set will
draw particle energies directly from this distribution, thereby
minimizing processing time. Venkataramaiah~{\it et al.} found this
approximation to be accurate within a one-percent margin of error, when compared to the
tabulated values of Rose~\cite{rose1955}. Because the interface of the $\mathrm{\beta}$-Fermi-function class is independent of the model implemented therein, it is easily possible for the user to substitute the RDM-extended's $\mathrm{\beta}$-Fermi-function approximation, by one which is
has better accuracy for a certain isotope.

\subsection{Decay Chains and Activation}
\label{sec:bateman}
Isotopes resulting from a radioactive decay are often unstable themselves. This leads to chains of subsequent decays until a stable daughter product is reached. Often it is desirable to only fully simulate those isotopes in a chain which are of interest as observables, either due to their half life or due to the radiation emitted when they decay. In such a case a full Monte-Carlo simulation of every decay is very inefficient. The sampling of interesting isotopes only requires determining the number of nuclei of a given species present in a chain at a specific time. 

Additionally, radioactive isotopes may be created by nuclear activation, often as a result of proton or neutron collisions with a nucleus. While the collisions and resulting activation is handled by the hadronic processes of Geant4~\cite{2006AIPC..867..479W}, bookkeeping of the activation buildup and the decay of the created unstable nuclei are tasks for the radioactive decay simulation. In the RDM-extended the activation simulation is also capable of altering the material composition on a per-volume basis to the isotope composition present at discrete user-defined time steps.

For a system of $n$ nuclei, where the $i$th nuclei decays into the
$(i+1)$th nuclei of the chain, this calculation can be done by solving
a system of coupled differential equations with the general form:

\begin{eqnarray}
\label{eqn:dgl2}
\frac{dN_{1}(t)}{dt} & = & -k_{1} N_{1}(t) \nonumber \\
\frac{dN_{2}(t)}{dt} & = & \lambda_{1}N_{1}(t)-k_{2} N_{2}(t) \nonumber \\
& \vdots & \nonumber \\
\frac{dN_{n}(t)}{dt} & = & \lambda_{n-1}N_{n-1}(t)-k_{n} N_{n}(t)
\end{eqnarray}
with $N_{i}$ being the quantity of the $i$th nucleus at time $t$ and $k_{i}=\lambda_{i}+\alpha_{i}$ containing information on the decay constant $\lambda_{i}$  and a constant particle number dependent activation rate $\alpha_{i}$.

A general approach for solving equation~(\ref{eqn:dgl2}) for any number of products was first derived by
Bateman~\cite{Bateman1910} and the above equations have as such become
known as the Bateman equations. 

One of the main disadvantages of the
original Bateman solution is that it does not consider branching,
i.e. if a parent nucleus can decay to different daughter nuclei via
different decay types. Many computational algorithms using Bateman's
approach or a derivative thereof exist, with the one implemented in
the original radioactive decay code~\cite{TruscottG4}, being just one
example which overcomes the branching limitation and includes
activation. These algorithms are generally computationally expensive,
in the sense that they require recursive loops for each nucleus in a
decay chain, but may achieve good overall performance, if the
calculations can be reused. 

An alternative is the algebraic approach
derived by M. Amaku {\it et al.} in~\cite{2010CoPhC.181...21A},
building upon work by R.~J.~Onega~\cite{onega:1019},
D.~Pressyanov~\cite{2002AmJPh..70..444P}, L.~Moral and
A.~F.~Pacheco~\cite{2003AmJPh..71..684M}, as well as
T.~M.~Senkov~\cite{2004AmJPh..72..410S} and D.~Yuan and
W.~Kernan~\cite{2007JAP...101i4907Y}. In this approach the properties
governing the decay chain are written into a matrix in Hessenberg-form
\begin{equation}
\Lambda =	\begin{bmatrix}
		-k_{1} & 0 & 0 & 0 \\
		k_{1} & -k_{2} & 0 & 0 \\
		0 & \ddots & \ddots & 0 \\
		0 & 0 & k_{n-1} & k_{n} \\
	\end{bmatrix}
\end{equation}
and a vector $\mathbf{N}(t)$ with
\begin{equation}
\mathbf{N}(t) = \begin{bmatrix}
		N_{1}(t) \\
		N_{2}(t) \\
		\vdots	 \\
		N_{i}(t) \\
		\end{bmatrix}.
\end{equation}
The system of equations~(\ref{eqn:dgl2}) may then be written as
\begin{equation}
 \frac{d\mathbf{N}}{dt} = \Lambda\mathbf{N}.
\end{equation}
This original algebraic approach was introduced by Onega in 1969 and
extended by Yuan and Kernan as well as Semkow to include branching by
modifying $\Lambda$ to
\begin{equation}
\Lambda = \begin{bmatrix}
	 \Lambda_{11} & & & & \\
	 \Lambda_{21} & \Lambda_{22} & & & \\
	 \vdots & \vdots & & \hspace{-1.5cm} \ddots & & \\	
	 \Lambda_{i1} & \Lambda_{i2} &\dots & \Lambda_{i,i} \\
	 \vdots & \vdots & & \vdots & \hspace{-1.cm}\ddots\\
	 \Lambda_{n1} & \Lambda_{n2} & \dots &\Lambda_{ni} & \Lambda_{nn} \\
	 \end{bmatrix}
\end{equation}
with 
\begin{equation}
\label{eqn:lambdBranch}
	\Lambda_{ij} = k_{i-1}b_{ij}
\end{equation}
for $i>j$ and 
\begin{equation}
	\Lambda_{ii} = -k_{i}.
\end{equation}
In equation~(\ref{eqn:lambdBranch}) $b_{ij}$ denotes the branching
ratio from the $j$th to the $i$th component of the decay chain with
$\sum^{n}_{i=j+1} b_{ij} = 1$. In computational practice $\Lambda$ can
be easily constructed by iterating through the decay chain and taking
activation rates into account, if necessary. Because $\Lambda$ is
independent of the actual nuclei numbers, it must only be constructed
once for each set of nuclei and activations characterizing a given
chain. Calculations using different nuclei numbers can reuse these
initial matrices as needed. Using the above definitions, the number of
nuclei of each species in the decay chain is then given by
\begin{equation}
 \mathbf{N}(t) = e^{\Lambda t}\mathbf{N}(0)
\end{equation}
which can be rewritten to 
\begin{equation}
	\label{eqn:sol}
	\mathbf{N}(t) = C\,e^{\Lambda_{d}t}\,C^{-1}
\end{equation}
as described by Onega. In equation~(\ref{eqn:sol}) $C$ is a square
matrix with the $n$th column consisting of the $n$th eigenvector of
$\Lambda$, so that $C =
[\mathbf{c}_{1},\mathbf{c}_{2},\dots,\mathbf{c}_{n}]$. $C^{-1}$ is its
inverse and $\Lambda_{d}$ a diagonal matrix with the elements
$\Lambda_{d,nn}$ being the $n$th eigenvalue of $\Lambda$.

M.~Amaku {\it et al.} then derive an algorithmic approach for calculating
the matrices $C$, $C^{-1}$ and $\Lambda_{d}$, which is computationally
less expensive and more accurate than a general approach of
numerically calculating the matrix elements. The elements of
$C=[c_{ij}]$ can be calculated with the recurrence expression
\begin{equation}
 \label{eqn:ccalc}
 c_{ij} = \frac{\sum^{i-1}_{k=j} \Lambda_{ik}c_{kj}}{\Lambda_{jj} - \Lambda_{ii}}
\end{equation} 
for $i=2,\dots,n$, $j=1,\dots,i-1$ and $c_{jj}=1$. Similarly the
elements of $C^{-1}=[c^{-1}_{ij}]$ are given by
\begin{equation}
 \label{eqn:cinvcalc}
 c^{-1}_{ij} = \sum^{i-1}_{k=j} c_{ik}b_{kj}
\end{equation} 
for $i>j$ and $b_{jj}=1$. Using $C$ and $C^{-1}$, $\Lambda_{d}$ is given by
\begin{equation}
 \label{eqn:lambda}
 \Lambda_{d} = C^{-1}\,\Lambda\,C
\end{equation} 
resulting in a general form for~\ref{eqn:sol} of
\begin{equation}
 \label{eqn:batemanfinal}
 \mathbf{N}(t) = C \begin{bmatrix}
  e^{-\Lambda_{d,11}t} & & & \\
  & e^{-\Lambda_{d,22}t} & & \\
  & & \ddots & \\
  & & & e^{-\Lambda_{d,nn}t} 
 \end{bmatrix} C^{-1} \mathbf{N}(0).
\end{equation}
In the RDM-extended code the replenishment and generation of unstable isotopes
through activation is calculated using the above equations on a per-volume level. In a first simulation run, a
bookkeeping class {\it G4ActivationBookkeeping} keeps track of all
activations occurring in a given volume and calculates the new
material and isotope composition of the volume at a given time. In a second run, the
material properties are then updated within the geometry, and the
volumes are set as radioactive background sources, with the calculated
activity via the {\it G4RadDecayVolumeBookkeeping} class. This
class samples the radioactive background emission, which should be
generated alongside a primary particle and pushes it on the event
stack. In this way minimal user intervention is required.

\section{Verification of Sampling Methods}
\label{sec:verification}

The radioactive code implementations have been verified for consistency with ENSDF data. Since ENSDF is established in the experimental community as an authoritative reference, comparison of Monte Carlo models against it provides valuable information for the users of these codes. The results presented here are chosen to highlight the physics-performance improvements of the statistical sampling in the RDM-extended package when compared to the per-decay sampling of the original Geant4 RDM code. Accordingly, the refactored per-decay code, which has been verified to produce equivalent results to the approach used in the original Geant4 RDM, is not detailed further.

In addition to the $\mathrm{\gamma}$-ray and conversion electron
emission directly given in ENSDF, the verification of intensities and
energies of fluorescence and Auger-emission with the Bambynek {\it et al.} data~\cite{1972RvMP...44..716B} has been included. 

\subsection{Simulated Data Production}
\label{sec:prep}
The Geant4 simulations consisted of $10^{6}$ decays of an unstable nucleus in an otherwise empty geometry. Any resulting unstable daughter nuclei were not decayed further. The kinetic energy of radiation and particles resulting from these decays was recorded separately for each radiation type ($\mathrm{\alpha}$, $\mathrm{\beta}$, $\mathrm{\gamma}$, non-$\mathrm{\beta}$ electrons) into two histograms: the first histogram ranging from $0\text{---}30\,\mathrm{keV}$ with a binning size of $0.05\,\mathrm{keV}$, the second histogram ranging from $0\text{---}30\,\mathrm{MeV}$ with a binning size of $0.2\,\mathrm{keV}$. The two binning schemes were chosen in order to properly distinguish discrete radiation in the X-ray and $\mathrm{\gamma}$--energy regime.

Using the above approach, data for $2910$ parent excitation levels of
isotopes were simulated using the original RDM code with radioactive
decay database version 3.3 and the per-decay simulation. The statistical
approach simulations included $3040$ parent isotopes and excitation
levels, i.e. $130$ more than the sample used for the production with per-decay sampling. The numbers differ due the different versions of ENSDF data and different parsers used for reprocessing. For both sets of simulations
$\mathrm{\beta}$-emission was distinguished from Auger- and conversion
electron emission.

\subsection{Evaluated Data Preparation}
The verification data were extracted from ENSDF into a tabulated form suitable for further analysis using the RADLIST~\cite{radlist} program. By using this ENSDF-supplied parser it was assured that no errors were introduced into the evaluated data during the automated extraction.

\subsection{Data analysis}
The automated, comparative analysis takes the level and particle energies from
the evaluated data as an initial energy estimate $E_{\mathrm{0,level}}$ or $E_{\mathrm{0,particle}}$ for where
emission may be present in the simulated data. The simulated data
histograms are then scanned in an interval of
$E_{0}\pm0.5\,\mathrm{keV}$ at energies below $30\,\mathrm{keV}$ and
$E_{0}\pm5.0\,\mathrm{keV}$ at higher energies for emission events. Data outside these windows was considered not to belong to the currently compared emission energy.

In case of discrete emission contained in a single bin the intensity
of the emission is given by
\begin{equation}
 \mathrm{Intensity} = \frac{\mathrm{Events\;in\;bin}}{\mathrm{Number\;of\;simulated\;decays}}.
\end{equation}
The energy uncertainty is
determined by the bin size. The intensity uncertainty is $\Delta{I}=\sqrt{{N}}/N_{\mathrm{sim}}$ with the event number ${N}$ per bin and the simulated number of decays $N_{\mathrm{sim}}$.

Should the emission be distributed into (multiple) neighboring bins,
which is possible due to recoil from previous emissions transferred
onto the emitting nucleus, the intensity is calculated from the total
number of the events in these bins. The evaluated data can indicate
that emissions at a different energy, but in proximity to the energy currently being processed may have also contributed
to the number of simulated events. In this case the summed intensities of all emissions in neighboring bins
are used for further comparisons. The energy position of these emissions is then set to
the median energy of the events, accordingly its uncertainty
is the median deviation.

\begin{figure}[!hb]
 \centerline{\includegraphics[width=3.5in]{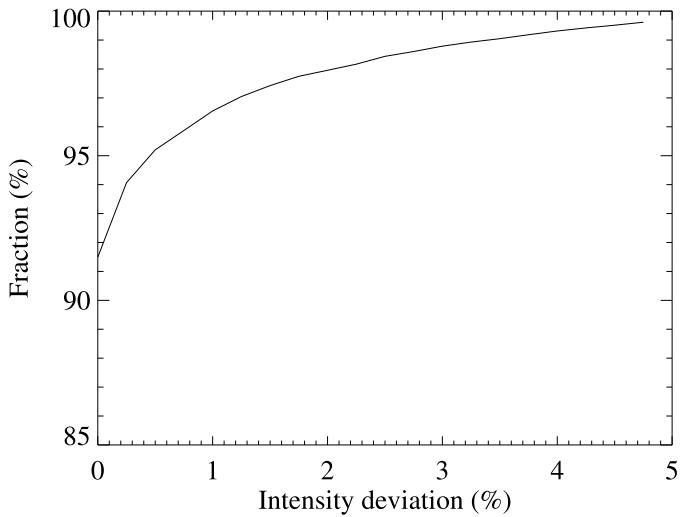}%
}
 \caption{Percentage of simulated emission within a given deviation from the simulation input.}
 \label{fig:idev_parser_comp}
\end{figure}

These steps provide energy and intensities of the
radiation from the evaluated and simulated data. Absolute
and relative energy and intensity deviations can then be calculated. In
this way overall comparisons (see Section~\ref{sec:gen_overview}) and
comparisons of individual isotopes, such as those measured
in~\cite{RadDecay2012_2}, are possible.

In order to assess the consistency of the simulation model, a complete
set of simulations using the statistical approach was
compared to the simulations' input data. Ideally, one would not expect any energy deviations at all and
the intensity deviations should be compatible with the statistical
error of the simulation. In practice the simulations' output is binned spectra, in order to keep total data amounts at a manageable
level. Due to this, in rare cases the emission at nearby energies may not
be distinguished properly, which in turn leads to intensity deviations. As is apparent from
Fig.~\ref{fig:idev_parser_comp}, $99\%$ of all simulated emissions are
within an intensity deviation of $3\%$ of the simulation input.

\section{Results: Consistency of the Radioactive Decay Codes with ENSDF}
\label{sec:gen_overview}
In the following we focus upon identifying global trends and
exploring the regions in which the radioactive decay simulations gives
reliable results. A validation with measured data can be found
in~\cite{RadDecay2012_2}.

\begin{figure}[htb]
\centering
\centerline{\includegraphics[width=3.5in]{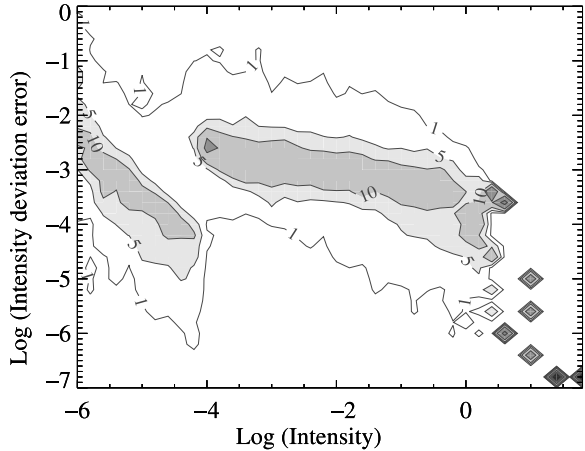}}
\caption{Distribution of the relative error of the observed deviations between simulated and evaluated radiation intensities with respect to the level intensity. The contour levels correspond to the percentile of values at given deviation error with respect to the number of values at a given level intensity. At higher intensities values are more sparse resulting in the observed "single box" contours.}
\label{fig:error_sum}
\end{figure}

For the comparisons presented in the following, knowledge of the comparisons' uncertainties is important. Fig.~\ref{fig:error_sum} shows a compilation of the intensity deviation uncertainties, which depend on the uncertainty of the evaluated intensity for a given level and the statistical uncertainty of the simulated data. As is apparent from the figure, even at low intensities the uncertainties are in a range below $10\%$ for $99\%$ of the data points.

\subsection{Intensity Deviations}

In order to help readers quickly identify the simulations' intensity discrepancy
for isotopes occurring in their simulation, it was chosen to display
the results as nuclide charts (a colored online version is available).

\begin{figure*}[!htbp]
\leftline{\hspace{2cm}\framebox{original RDM}
\hfil
\hspace{2cm}\framebox{RDM-extended}}
\vspace{0.2in}
\centerline{{\includegraphics[width=3.5in]{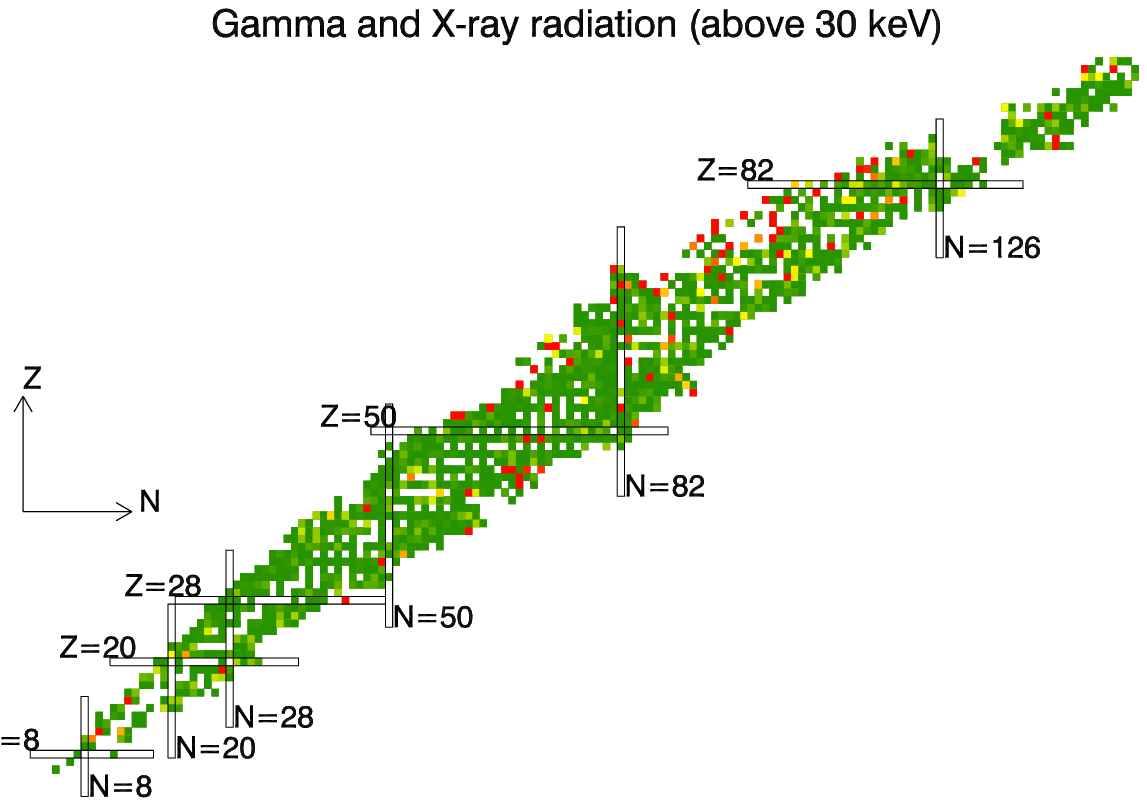}}
\hfil
{\includegraphics[width=3.5in]{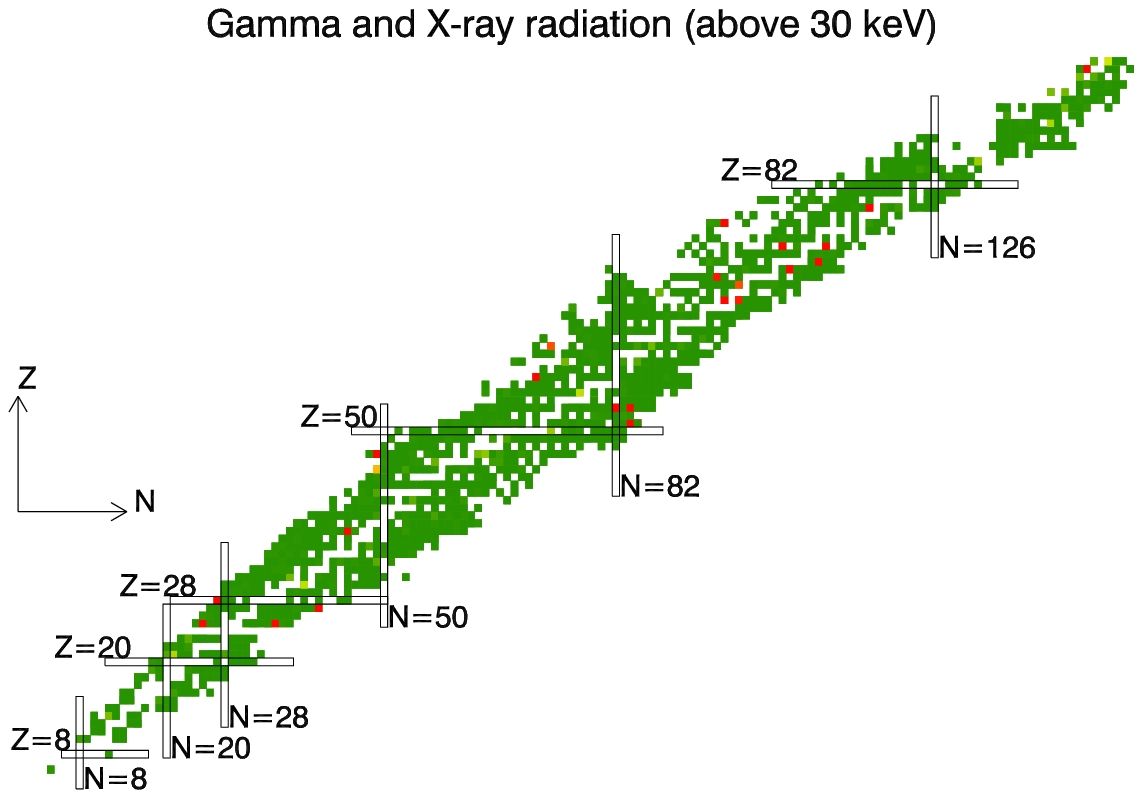}
}}
\vspace{0.2in}
\centerline{{\includegraphics[width=3.5in]{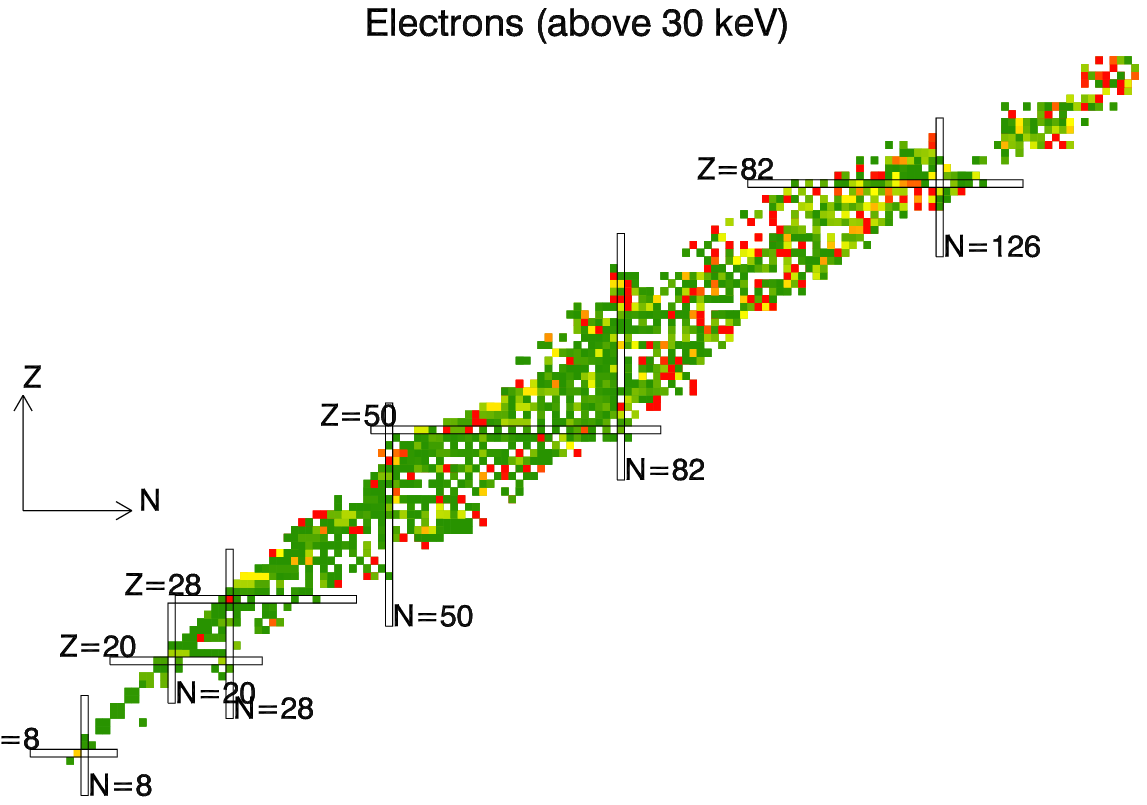}}
\hfil
{\includegraphics[width=3.5in]{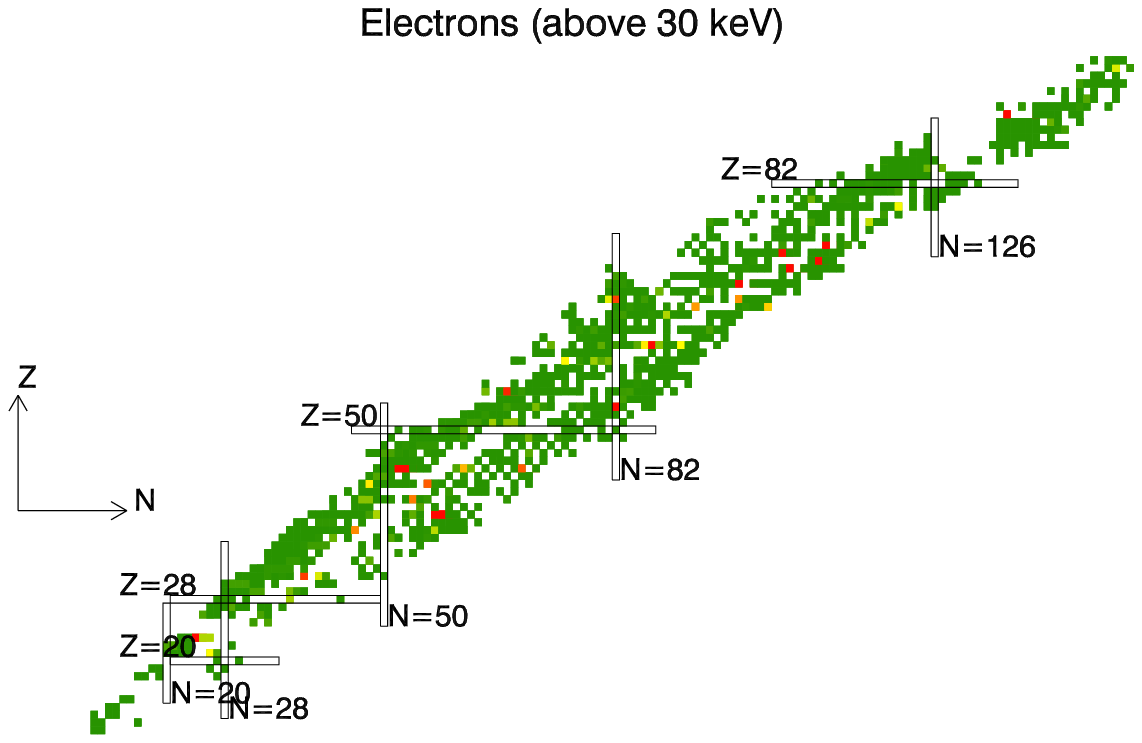}
}}
\vspace{0.2in}
\centerline{{\includegraphics[width=3.5in]{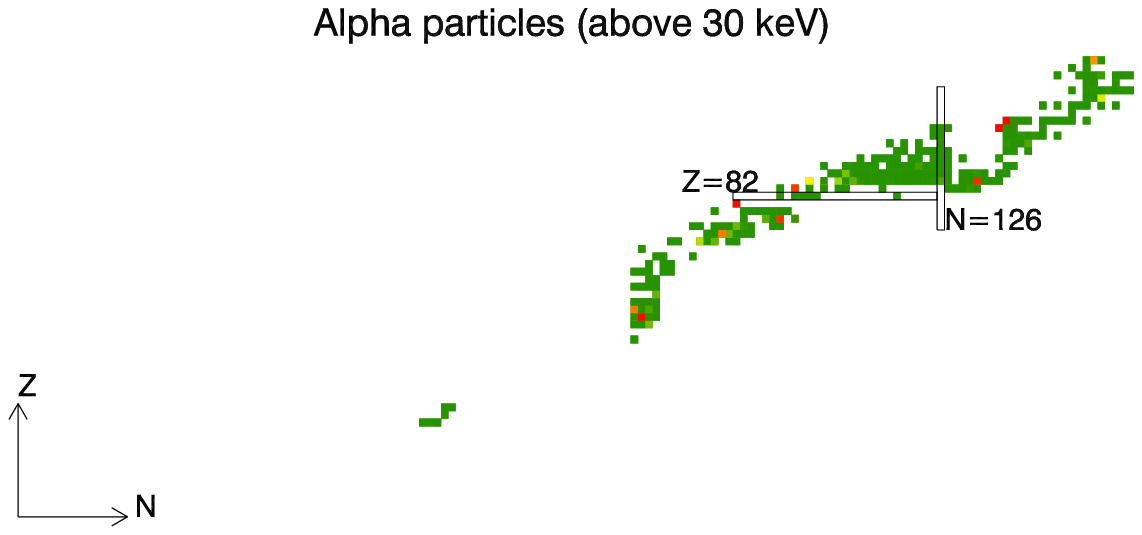}}
\hfil
{\includegraphics[width=3.5in]{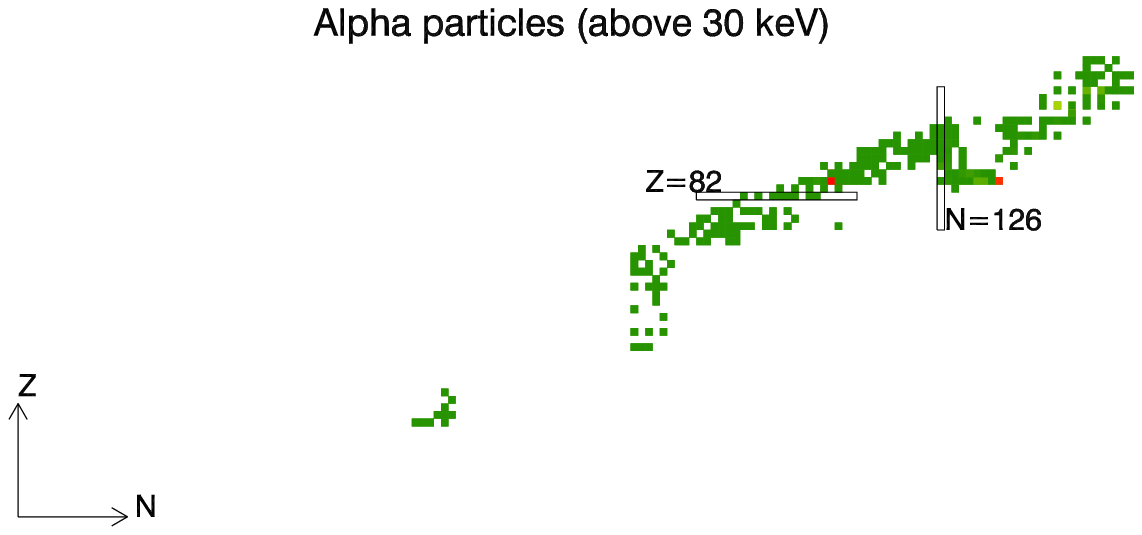}
}}
\vspace{0.4in}
\centerline{\includegraphics[width=3.5in]{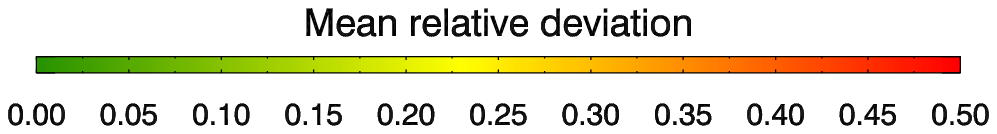}}
\vspace{-0.3in}
\caption{Nuclide charts showing the median relative intensity
 deviations per isotope for $\mathrm{\gamma}$- (top),
 conversion-electron (middle) and $\mathrm{\alpha}$-emission
 (bottom). Simulations using the per-decay approach of the original Geant4 RDM
 are shown on the left; simulations using the RDM-extended statistical sampling method
 are shown on the right.}
\label{fig:nuclide_high_energy}
\end{figure*}

Fig.~\ref{fig:nuclide_high_energy} shows a comparison of the
intensity discrepancy of the original Geant4 RDM per-decay sampling alongside the discrepancy of
the statistical approach of the RDM-extended when compared to ENSDF data. As is
apparent from the figures, both sampling methods reproduce the
$\mathrm{\gamma}$- and $\mathrm{\alpha}$-emission intensities given in
ENSDF within a few percent deviation for the majority of isotopes. Specifically, the mean deviations
$(I_{\mathrm{exp}}-I_{\mathrm{sim}})/I_{\mathrm{exp}}$ amount to
$(8.57\pm2.22)\% $ for $\mathrm{\gamma}$-rays and $(2.47\pm2.26)\% $
for $\mathrm{\alpha}$-emission when using the per-decay code. 

For the
statistical approach the deviations are minimal, with $(1.85\pm2.07)\%
$ for $\mathrm{\gamma}$-rays and $(5.61\pm2.62)\% $ for
$\mathrm{\alpha}$-emission. The outliers ($>50\%$ discrepancy) in the simulation using the
RDM-extended package can be explained by the way the ENSDF data are
parsed into the radioactive data library. In cases where multiple
datasets describing the same emission exist, the implemented parser
is tuned to pick experimentally determined data or, if such a
distinction is not possible, use the first data set. The RADLIST
program preferably uses theoretical data according to the
documentation~\cite{radlist}.

It is further apparent from Fig.~\ref{fig:nuclide_high_energy} that the original
per-decay code does not reproduce the ENSDF intensities of
conversion electrons well. This manifests itself in a mean deviation of
$(35.67\pm6.32)\%$ compared to $(9.54\pm0.98)\%$ for the statistical
approach.

For the original Geant4 RDM per-decay code the $\mathrm{\gamma}$- and conversion electron
intensity deviations must be attributed to the {\it
 G4PhotonEvaporation} model or more specifically its underlying data
library~\cite{Pia4237413}. A comprehensive verification and validation of this process
would be beyond the scope of this paper and was thus not undertaken.

\begin{figure*}[!h]
\leftline{\hspace{2cm}\framebox{original RDM}
\hfil
\hspace{2cm}\framebox{RDM-extended}}
\vspace{0.2in}
\centerline{{\includegraphics[width=3.5in]{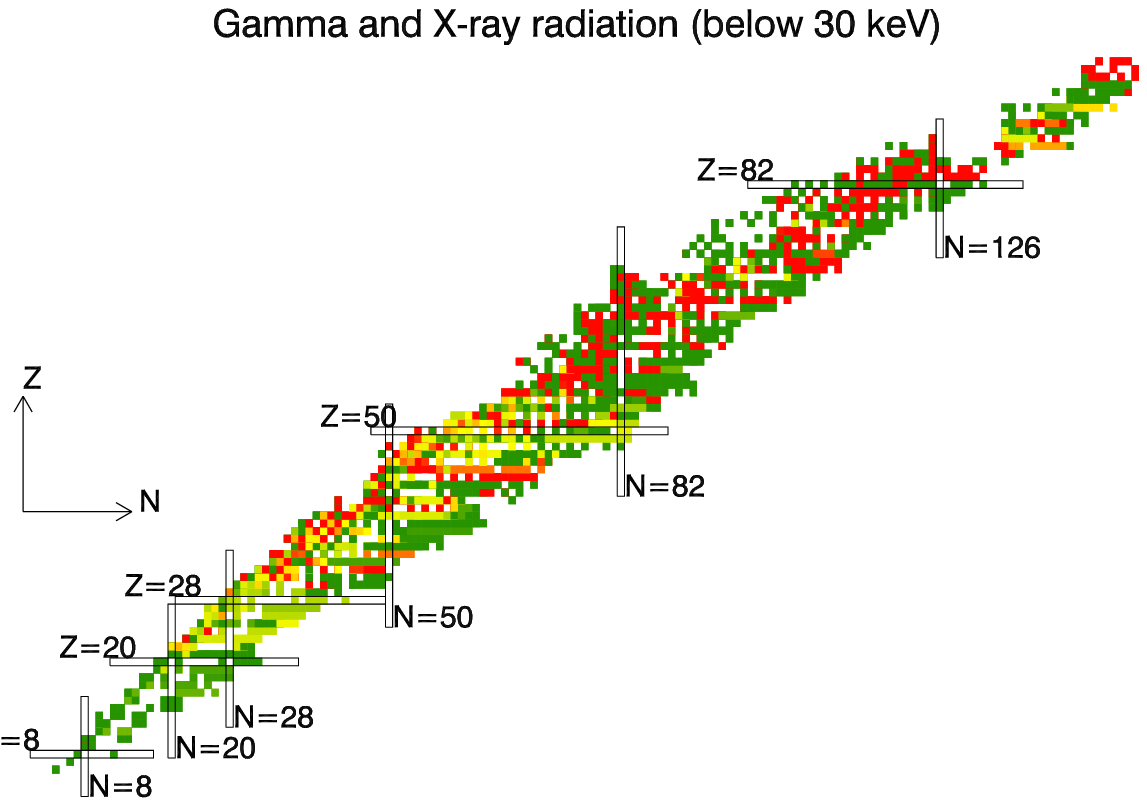}}
\hfil
{\includegraphics[width=3.5in]{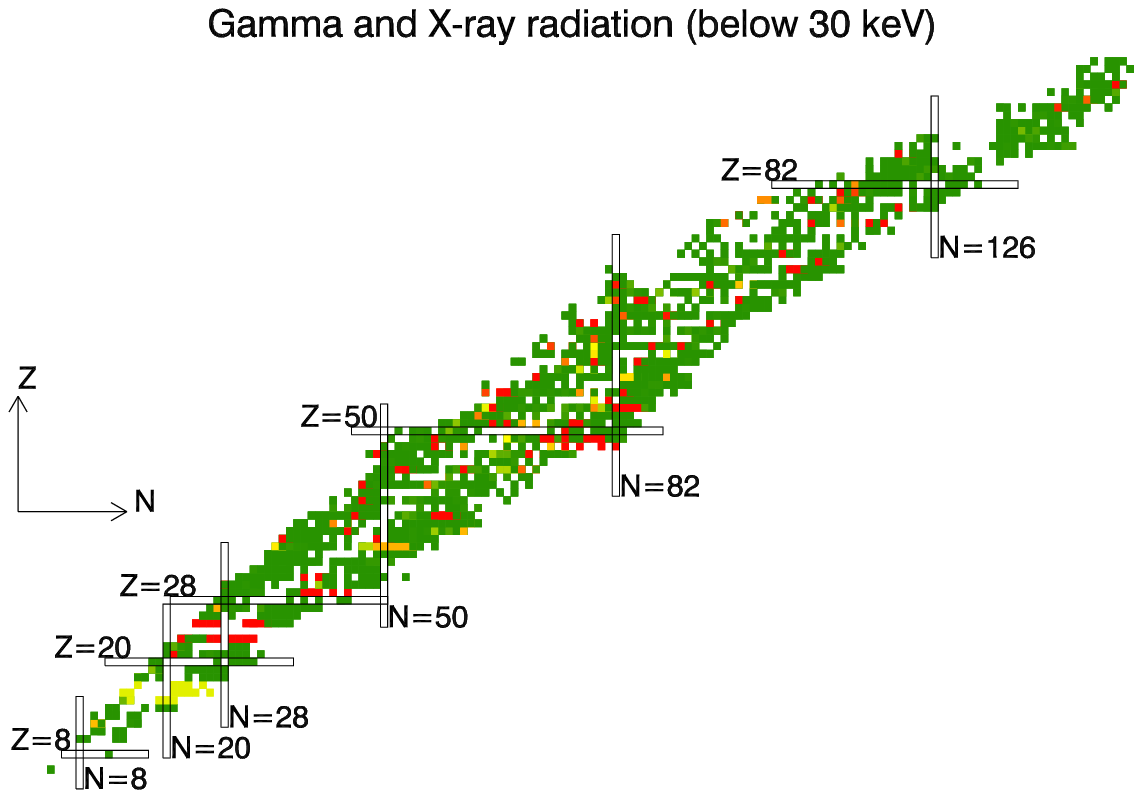}
}}
\vspace{0.2in}
\centerline{{\includegraphics[width=3.5in]{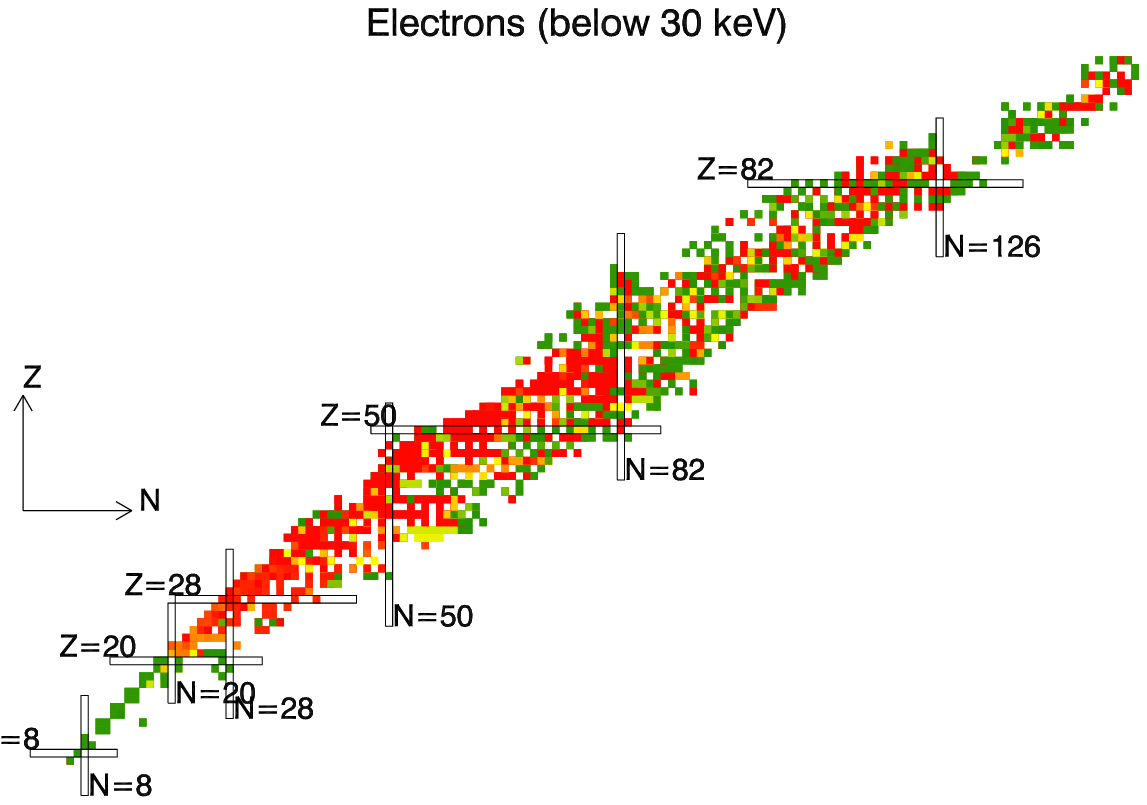}}
\hfil
{\includegraphics[width=3.5in]{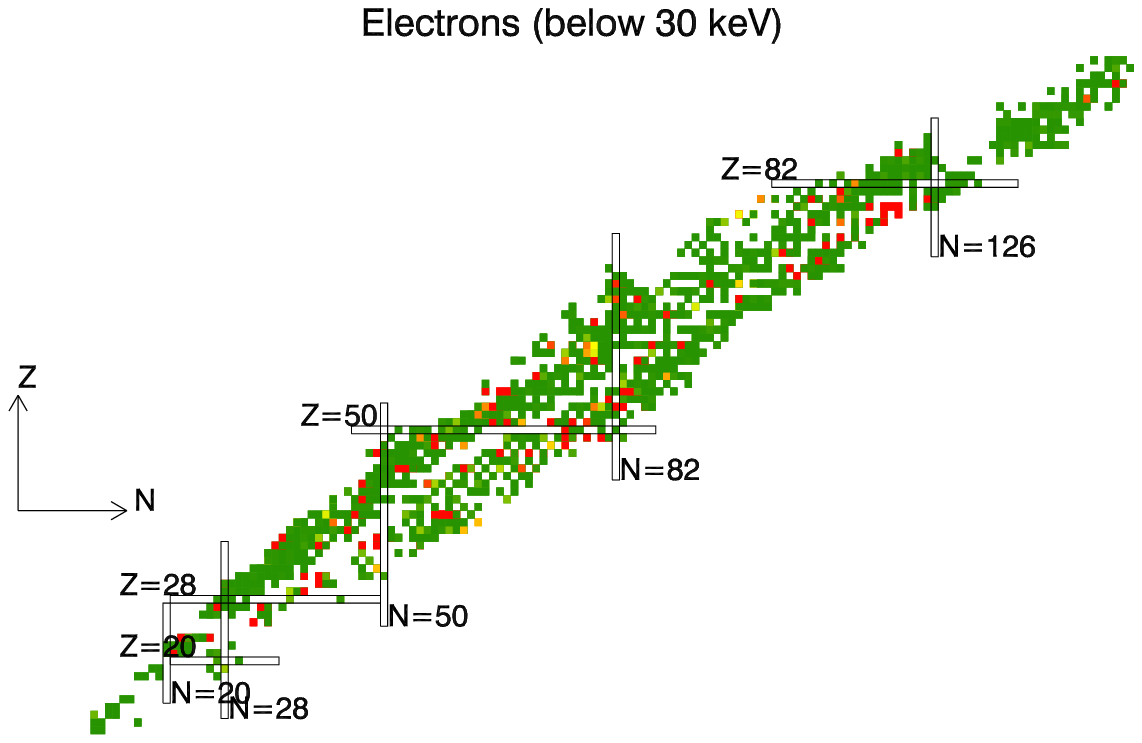}
}}
\vspace{0.4in}
\centerline{\includegraphics[width=3.5in]{hauf13}}
\vspace{-0.3in}
\caption{Nuclide charts showing the median relative intensity
 deviations per isotope for X-ray (top) and Auger-electron emission
 (bottom). Simulations using the per-decay approach of the original Geant4 RDM
 are shown on the left; simulations using the RDM-extended statistical sampling method
 are shown on the right.}
\label{fig:nuclide_low_energy}
\end{figure*}

For fluorescence and Auger emission the deviation between evaluated
ENSDF intensities and those produced by the original RDM per-decay approach are even
larger, as is apparent from Fig.~\ref{fig:nuclide_low_energy}. For
X-ray emissions the deviation amounts to $(52.64\pm1.97)\%$; for
Auger-electron to $(52.57\pm0.96)\%$. Again not the radioactive decay
code alone is responsible for these offsets but its interplay with the
{\it G4AtomicDeexcitation} class and its associated EADL data library. It is
interesting to note that the deviations are largest for the isotope on
the left of the nuclide chart, i.e. isotopes which decay via electron
capture. This substantiates the conclusion that the EADL data library is the source of the deviations,
as it is only called by the per-decay approach if electron capture
decays need to be sampled (in Geant4 9.4, changed in Geant4 9.5). Again the extended-RDM's statistical method yields results with smaller deviations, which amount to $(4.09\pm1.78)\%$ for X-ray
emission and $(1.34\pm1.16)\%$ for Auger electron emission. This is a
more than 10-fold improvement in intensity consistency with respect to the per-decay approach of the original Geant4 RDM. A similar magnitude of deviations is reported in~\cite{pia2011evaluation} for EADL data, in which different binding energy data libraries were compared with reference data. In this work EADL consistently showed the largest deviations.

It should be stressed that the observed intensity deviations for the
per-decay approach result mainly from the incoherence of the data
libraries involved in the sampling of nuclear deexcitation and
fluorescence emission with the ENSDF-database, from which the
reference data were derived. For the statistical approach the
deviations are naturally smaller, as all radioactive decay data are
derived from a single ENSDF-library and supplementary
atomic data files.

\subsection{Energy deviations}
\begin{figure*}[!htbp]
\centering
\leftline{\hspace{2cm}\framebox{original RDM}
\hfil
\hspace{2cm}\framebox{RDM-extended}}
\vspace{-0.3cm}
\centerline{\subfloat{\includegraphics[width=3.5in]{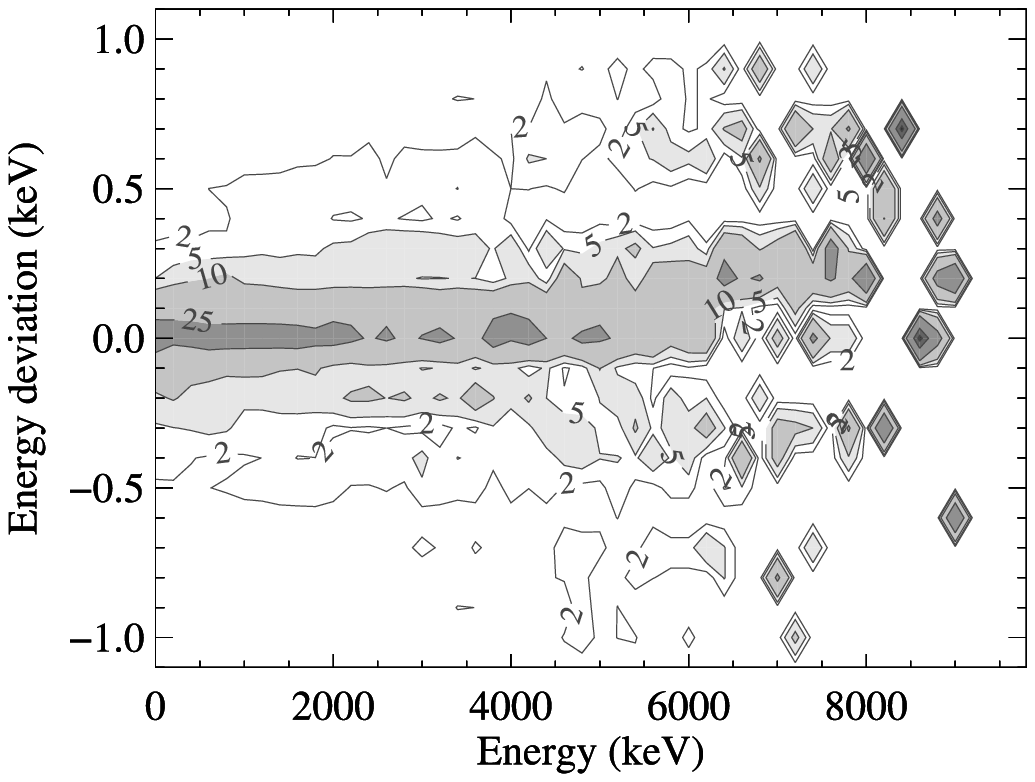}}
\hfil
\subfloat{\includegraphics[width=3.5in]{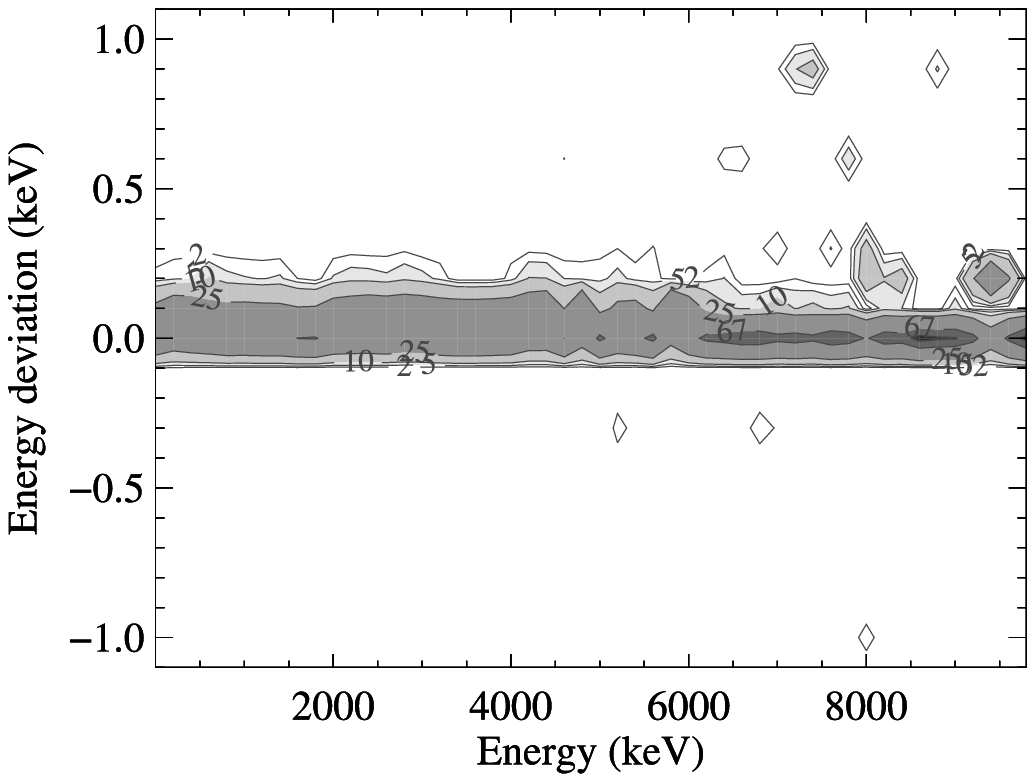}
}}
\caption{Distribution of absolute energy deviations with respect to the
 radiation energy for all level energies. The contour levels
 correspond to the percentile of values at a given energy deviation
 with respect to the number of values at a given radiation energy.
 At higher energies evaluated values are more sparse resulting in
 fractured contours. Simulations using the per-decay approach of
 the original Geant4 RDM are shown on the left; simulations using the
 RDM-extended statistical approach are shown on the right. }
\label{fig:energy_energy}
\end{figure*}

\begin{figure*}[!htbp]
\centering
\leftline{\hspace{2cm}\framebox{original RDM}
\hfil
\hspace{2cm}\framebox{RDM-extended}}
\vspace{-0.3cm}
\centerline{\subfloat{\includegraphics[width=3.5in]{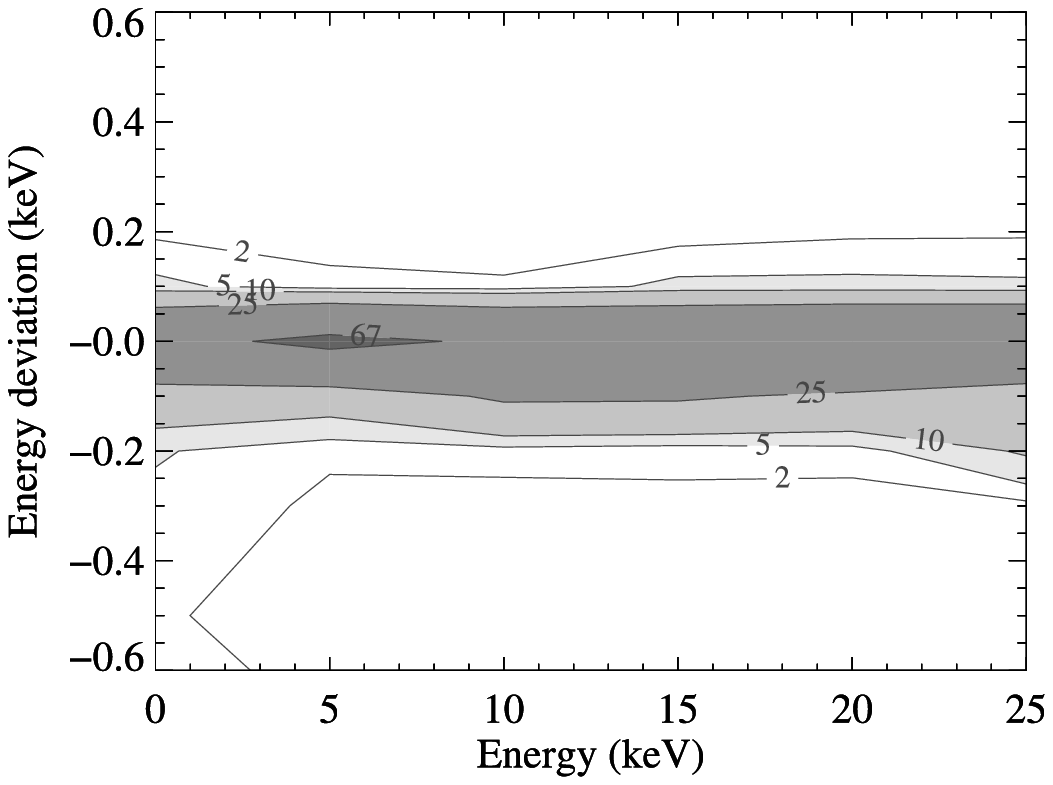}}
\hfil
\subfloat{\includegraphics[width=3.5in]{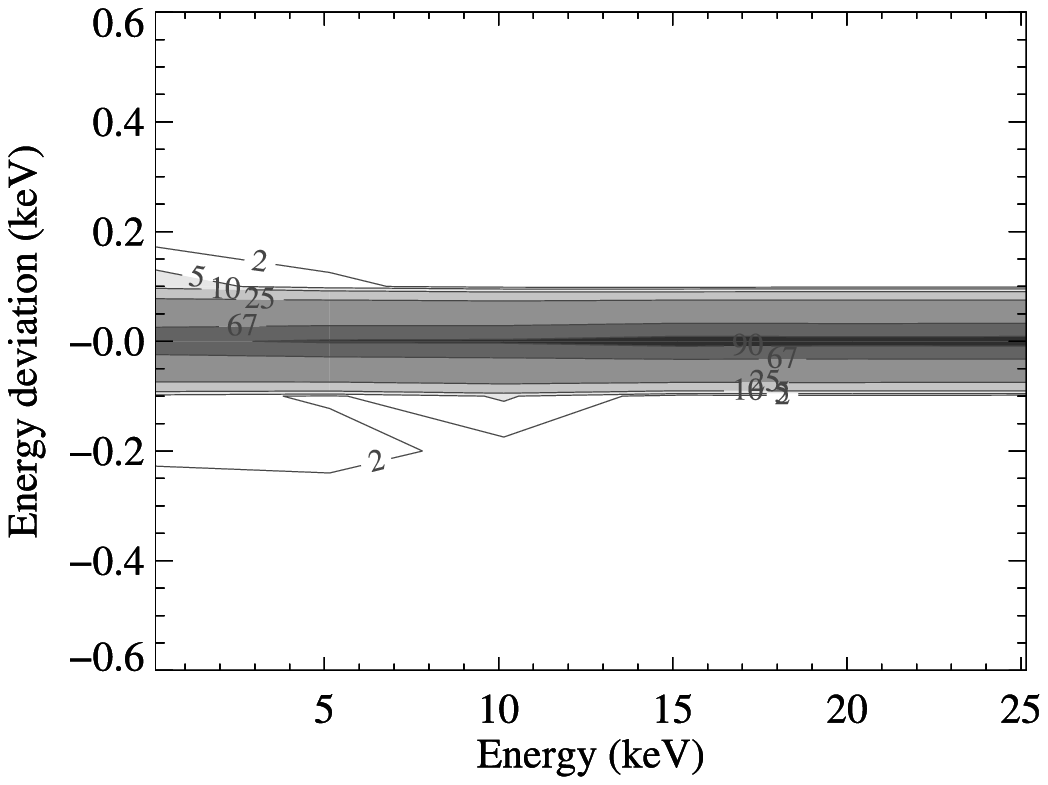}
}}
\caption{Distribution of absolute energy deviation with respect to the
 radiation energy for level energies below $25\,\mathrm{keV}$. The
 contour levels correspond to the percentile of values at a given
 energy deviation with respect to the number of values at a given
 radiation energy. Simulations using the per-decay approach of
 the original Geant4 RDM are shown on the left; simulations using the
RDM-extended statistical approach are shown on the right. }
\label{fig:energy_low}
\end{figure*}

Most application scenarios will depend on the correct sampling of the
energy of the $\mathrm{\gamma}$- or X-ray radiation and that $\mathrm{\alpha}$
particles as well as Auger- and conversion-electrons are emitted at
the experimentally determined energies. From the comparisons shown in
Fig.~\ref{fig:energy_energy} and Fig.~\ref{fig:energy_low}, one can
conclude that at energies in the X-ray and Auger-electron range less than
$30\,\mathrm{keV}$, a deviation of less than $0.2\,\mathrm{keV}$ with respect to ENSDF data is to be
expected for all radiation, when using the original Geant4 RDM per-decay
method. Statistical sampling is again more consistent with 
ENSDF data. Here the observed energy deviation is less than
$0.1\,\mathrm{keV}$ for the majority of emissions. For
$\mathrm{\gamma}$-radiation, $\mathrm{\alpha}$-particles and
conversion electrons at higher energies, deviations of less than
$0.5\,\mathrm{keV}$ are observed with the per-decay approach. The
statistical approach shows deviations at or less than the bin size of
$0.2\,\mathrm{keV}$.

\section{Computational Performance}
\label{sec:perf}
\begin{figure}[!htbp]
 \centerline{\includegraphics[width=3.in]{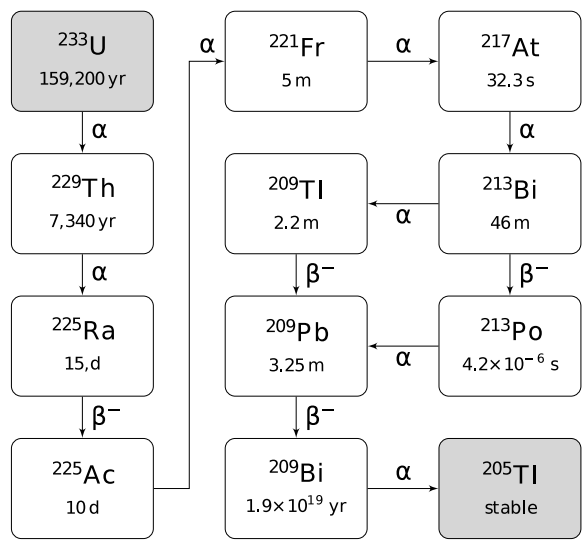}%
 }
 \caption{The $\mathrm{^{233}U}$ decay chain simulated for
  performance testing. The initial isotope of the chain was varied
  from $\mathrm{^{233}U}$ to $\mathrm{^{213}Bi}$ to test different
  chain lengths.}
 \label{fig:decaychain}
\end{figure}

Computational performance is a significant aspect for large scale
Monte-Carlo simulations. The RDM-extended package was
designed keeping this in mind (Section~\ref{sec:bateman}). The implementation reflects this performance optimization: for instance, it uses hash-maps ({\it
 std::unordered\_map}) from the {\it C++11}-standard~\cite{cpp11}, which allow
element access times which scale with the number of elements in $O(1)$, instead of $O(log(n))$, as would be the
case for traditional {\it std::map} containers (see e.g. \cite{2010arXiv1012.3292H}). In order to estimate
the performance gain achievable using this implementation in
comparison to the original RDM, the following performance tests were
undertaken:
\begin{itemize}
\item Decay an isotope $100\,000$ times using the refactored per-decay and the
 statistical approach.
\item Decay a chain of isotopes and retrieve all emission which has
 occurred from all isotopes in the chain within a sampled time period of $\Delta
 t$, which could be, for instance, an experiment's measurement duration, $10\,000$ times.
\item Decay a chain of isotopes and retrieve the emission from a
 single isotope within the chain occurring within a time period
 $\Delta t$ $10\,000$ times.
\end{itemize}
Each of the above tests was repeated $200$ times to ensure that
temporary CPU load from the operating system or other processes would
not bias the results. For the first test the isotopes
$\mathrm{^{22}Na}$($\mathrm{\beta^{+}}$),
$\mathrm{^{60}Co}$($\mathrm{\beta^{-}}$),
$\mathrm{^{229}Th}$($\mathrm{\alpha}$) and $\mathrm{^{133}Ba}$(EC)
were decayed, constituting an example for each of the particle emitting decay
types. As an example of a full decay chain, the
$\mathrm{^{233}U}$ decay chain shown in
Fig.~\ref{fig:decaychain} was simulated. In order to estimate the performance for
different decay chain lengths, different initial nuclei in this chain were chosen. 

All performance tests were done on a $12$-core {\it XEON} machine at
$2.93\,\mathrm{GHz}$ running Ubuntu 10.10 {\it Maverick Meerkat}
and an identical application based on {Geant4 9.4p04}, which was unaltered except for the decay code. The
{\it gcc 4.4.5} compiler with Geant4 standard compiler flags and the
extensions of the {\it C++11}-standard enabled was used for
compilation. The radioactive decay was the only physics simulation process
included in the test environment, thus guaranteeing that no other
processes, which the involved particles might be subject to during further
tracking, would influence the results.

\begin{figure}[!htbp]
\centerline{\includegraphics[width=3.5in]{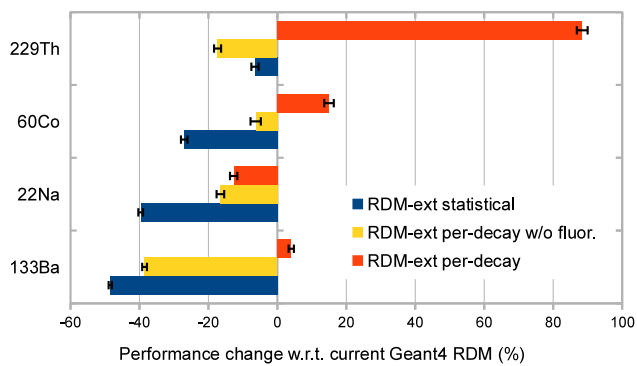}%
}
\caption{Computational performance change when using the RDM-extended package with the statistical approach in comparison to the original Geant4 RDM and the per-decay approach. Each simulation consisted of $100\,000$ decays and was repeated $200$ times. Three different sampling techniques are shown: statistical approach, classical approach and classical approach without post-decay fluorescence production. This last approach resembles the original Geant4 RDM implementation.}
\label{fig:performance_1}
\end{figure}
 
The relative performance of the RDM-extended statistical sampling method
shown in Fig.~\ref{fig:performance_1} asserts that this method is -- in almost all cases more than $20\,\mathrm{\%}$ -- faster
than the per-decay approach implemented in the original Geant4 RDM. An
exception are isotopes which have a large number of deexcitation
emissions such as $\mathrm{^{229}Th}$. Here the linear increase of
sampling time with the number of emissions results in a performance
penalty.

When comparing the refactored per-decay approach decay emission production,
i.e. delegation to {\it G4PhotonEvaporation} and {\it
 G4AtomicDeexcitation}, two scenarios have been distinguished. In the
first case, labeled "per-decay" in
Fig.~\ref{fig:performance_1}, the vacant shell index, which may be
output by the {\it G4PhotonEvaporation} class at the end of
deexcitation, is passed to {\it G4AtomicDeexcitation} for
X-ray and Auger-electron production. The original Geant4 RDM code does not
pass the atomic shell index, a scenario also simulated with the RDM-extended package and labeled "per-decay w/o fluor." in
Fig.~\ref{fig:performance_1}. It is apparent from the figure,
especially for $\mathrm{^{229}Th}$, that the inclusion of {\it
 G4AtomicDeexcitation} results in a severe performance penalty. If a
full treatment of X-ray and Auger-electron emission is needed and
simulation performance is critical, it is thus recommended to use
statistical sampling, if the applications scenario permits it.

\begin{figure}[!htbp]
\centerline{\includegraphics[width=3.5in]{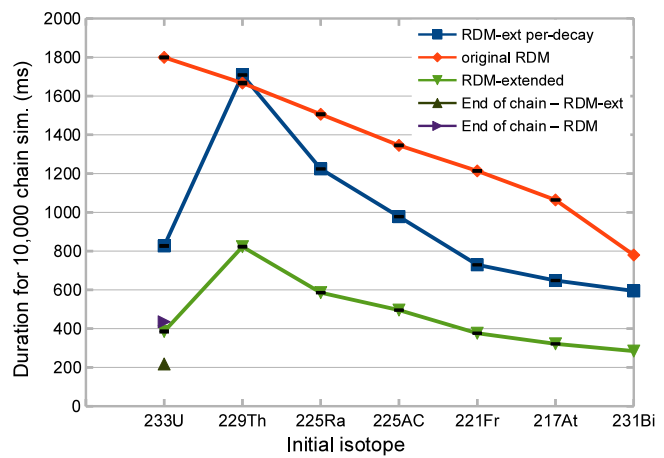}%
}
\caption{Computational performance of the RDM-extended package and the original RDM when decaying the $\mathrm{^{233}U}$ decay chain. The chain length was varied by setting different initial nuclei. The two data points labeled end of chain are the performance values for when only the emission of the last isotope in the chain is of interest.}
\label{fig:performance_2}
\end{figure}

For the decay chain performance comparison shown in
Fig.~\ref{fig:performance_2} one has to consider that the per-decay
approach of the original Geant4 RDM does not take the duration of the time period to be sampled into account. Instead it is up to the user to filter for
the relevant emission according to the application scenario. This explains the increase in computing
time needed by the RDM-extended between $\mathrm{^{233}U}$ as
the initial isotope and $\mathrm{^{229}Th}$. As is shown in
Fig.~\ref{fig:decaychain}, thorium has a much shorter half life time than uranium. Regardless of which isotope is taken as the initial one in the chain,
the simulated time duration for which decays are sampled stays the same at $3\times10^{13}\,\mathrm{s} =
95120\,\mathrm{yr}$. During this time, much less uranium than thorium will have decayed. Accordingly, when uranium is chosen as an initial isotope, less emission has to be sampled,
than is the case for thorium. This reflects itself in the performance increase of $\sim 55\,\mathrm{\%}$.

Similarly, the statistical sampling in the RDM-extended package is faster, if only the
emission from selected isotopes in the chain is of interest. In the RDM-extended code only the emissions from these isotopes are actually sampled and passed to tracking, reducing the number of particles which need to be simulated.
In the original RDM the user has to filter for the emission of interest after it has been produced by the decay code and has been passed to tracking. This is inefficient and
computing time intensive, as can be seen from Fig.~\ref{fig:performance_2}. An extreme case for such a scenario is when only the emission of the final isotope in a chain is of interest. This case is documented in Fig.~\ref{fig:performance_2} by the two ''end of chain'' data values.

\section{Conclusion}

Experimental requirements concerning radioactive decay simulation have been analyzed and evaluated against radioactive decay models and functionality available in present Monte-Carlo codes. It was found that none of the available codes offers the possibility to correctly simulate physics on the per-decay level and to correctly simulate the statistical outcome of many decays without unnecessary overhead within one framework. 

A software package, which addresses these requirements and allows simulations using both approaches, has been designed, implemented and verified with respect to the established reference of the ENSDF evaluated data library. The RDM-extended package described in this paper reproduces the functionality of the original Geant4 RDM package implementing per-decay sampling, although with improved software design and generally faster computational performance. In addition, it encompasses functionality for statistical sampling of radioactive decays: with respect to the per-decay approach of the pre-existing Geant4 RDM, this approach has been verified to achieve better consistency with ENSDF data, and better computational performance. Significant consistency improvements have been verified especially in the X-ray regime.

The RDM-extended package can be used transparently in Geant4-based simulation applications. Its experimental validation is documented in a distinct dedicated paper~\cite{RadDecay2012_2}.

\section*{Acknowledgment}

The authors would like to acknowledge the financial support from the Deutsche Zentrum fuer Luft-- und Raumfahrt (DLR) under Grant number 50QR902 and 50Q1102.




\bibliographystyle{IEEEtran}
\bibliography{IEEEabrv,all}
\end{document}